\documentclass[conference]{IEEEtran}
\usepackage{cite}
\usepackage{amsmath,amssymb,amsfonts}
\usepackage{algorithmic}
\usepackage{graphicx}
\usepackage{textcomp}
\usepackage{xcolor}
\usepackage{enumitem}
\usepackage[font={small}]{caption}
\usepackage{wrapfig}
\usepackage{balance}

\def\BibTeX{{\rm B\kern-.05em{\sc i\kern-.025em b}\kern-.08em
    T\kern-.1667em\lower.7ex\hbox{E}\kern-.125emX}}
 
\usepackage[T1]{fontenc}
\usepackage{mathtools}
\usepackage{enumitem,comment,wasysym}
\usepackage{color, soul}
\usepackage{pbox}
\usepackage{upquote}
\usepackage{listings}
\usepackage{tikz,subfigure}
\usepackage{url}
\usepackage{booktabs}
\usepackage{xspace}
\usepackage{hyperref}

\usepackage{breakurl}
\usepackage{comment}

\newcommand{\point}[1]{\vspace{.05in} \par\noindent\textbf{#1}. }

\definecolor{dkgreen}{rgb}{0,0.6,0}
\definecolor{gray}{rgb}{0.5,0.5,0.5}
\definecolor{mauve}{rgb}{0.58,0,0.82}
\definecolor{editorGray}{rgb}{0.95, 0.95, 0.95}
\definecolor{editorOcher}{rgb}{1, 0.5, 0}
\definecolor{editorGreen}{rgb}{0,0.6,0}
\definecolor{darkgreen}{RGB}{0,90,90}
\definecolor{lightgray}{rgb}{0.95, 0.95, 0.95}
\definecolor{darkgray}{rgb}{0.4, 0.4, 0.4}
\definecolor{orange}{rgb}{1,0.45,0.13}		
\definecolor{olive}{rgb}{0.17,0.59,0.20}
\definecolor{brown}{rgb}{0.69,0.31,0.31}
\definecolor{purple}{rgb}{0.38,0.18,0.81}
\definecolor{lightblue}{rgb}{0.1,0.57,0.7}
\definecolor{lightred}{rgb}{1,0.4,0.5}
\definecolor{dkgreen}{rgb}{0,0.6,0}
\definecolor{gray}{rgb}{0.5,0.5,0.5}
\definecolor{mauve}{rgb}{0.58,0,0.82}
\definecolor{editorGray}{rgb}{0.95, 0.95, 0.95}
\definecolor{editorOcher}{rgb}{1, 0.5, 0}
\definecolor{editorGreen}{rgb}{0,0.6,0}

\definecolor{MyMaroon}{HTML}{D50000}
\definecolor{MyYellow}{HTML}{FFD600}
\definecolor{MyGreen}{HTML}{00C853}

\definecolor{MyTeal}{HTML}{009688}
\definecolor{MyOrange}{HTML}{FF5722}
\definecolor{MyGray}{HTML}{E0E0E0}
\definecolor{MyIndigo}{HTML}{3F51B5}
\definecolor{MyPurple}{HTML}{9C27B0}
\definecolor{MyBlack}{HTML}{000000}
\definecolor{MyBlue}{HTML}{3F51B5}
\lstdefinelanguage{JavaScript}{
  morekeywords={typeof, window, new, true, false, catch, function, return, null, catch, switch, var, if, in, else, case, break, document, write, createElement, width, height, display, visibility, border, document.write},
  morecomment=[s]{/*}{*/},
  morecomment=[l]//,
  morestring=[b]",
  morestring=[b]'
}

\lstdefinelanguage{HTML5}{
  language=html,
  sensitive=true,	
  alsoletter={<>=-},	
  morecomment=[s]{<!-}{-->},
  tag=[s],
  otherkeywords={
  >,
	<!DOCTYPE,
  </html, <html, <head, <title, </title, <style, </style, <link, </head, <meta, />,
	</body, <body,
	</div, <div, </div>,
	</p, <p, </p>, <a, <h1, </h1>, </h1,
	</script, <script,
  <canvas, /canvas>, <svg, <rect, <animateTransform, </rect>, </svg>, <video, <source, <iframe, </iframe>, </video>, <image, </image>, <header, </header, <article, </article, </iframe, <img, <span, </span, </a,<button, </button>, <ul>, </ul>, <li>, </li>
  },
  ndkeywords={
  =, ===, ==,
  charset=, src=, id=, width=, height=, style=, type=, rel=, href=, name=, tabindex=, align=, scrolling=, framespacing=, frameborder=, allowtransparency=, data-dm-title=, data-dm-format=, data-dm-filesize=, target=, data-dm=, data-dm-icon=, data-dm-href-free=, data-dm-filename=, data-dm-hosted-file=, data-dm-href= , data-dm-carregado=, class=, alt=,
  fill=, attributeName=, begin=, dur=, from=, to=, poster=, controls=, x=, y=, repeatCount=, xlink:href=,
  margin:, padding:, background-image:, border:, top:, left:, position:, width:, height:, margin-top:, margin-bottom:, font-size:, line-height:,
  transform:, -moz-transform:, -webkit-transform:,
  animation:, -webkit-animation:,
  transition:,  transition-duration:, transition-property:, transition-timing-function:,
  }
}

\lstdefinestyle{htmlcssjs} {%
  backgroundcolor=\color{editorGray},
  basicstyle=\fontsize{8}{8}\ttfamily,
  frame=tb,
  captionpos=b,
  belowcaptionskip=\medskipamount,
  xleftmargin={0.5cm},
  numbers=left,
  stepnumber=1,
  firstnumber=1,
  numberfirstline=true,	
  identifierstyle=\color{black},
  keywordstyle=\color{blue}\ttfamily,
  ndkeywordstyle=\color{editorGreen}\ttfamily,
  stringstyle=\color{black}\ttfamily,
  commentstyle=\color{brown}\ttfamily,
  language=HTML5,
  alsolanguage=JavaScript,
  alsodigit={.:;},	
  tabsize=2,
  showtabs=false,
  showspaces=false,
  showstringspaces=false,
  extendedchars=true,
  breaklines=true,
  numberstyle=\tiny\color{gray},
  literate=%
  {Ö}{{\"O}}1
  {Ä}{{\"A}}1
  {Ü}{{\"U}}1
  {ß}{{\ss}}1
  {ü}{{\"u}}1
  {ä}{{\"a}}1
  {ö}{{\"o}}1
}
\newcommand{\AdLabel}{\textsf{AD}\xspace}
\newcommand{\NoAdLabel}{\textsf{NON-AD}\xspace}

\newcommand{\tool}{\textsc{AdGraph}\xspace}
\newcommand{\tools}{\textsc{AdGraph's}\xspace}

\newcommand{\JS}{JavaScript\xspace}
\newcommand{\DOM}{DOM\xspace}
\newcommand{\VJS}{V8\xspace}

\newcommand{\TP}{third party\xspace}

\newcommand{\AJAX}{\texttt{AJAX}\xspace}

\newcommand{\NumFeatures}{64\xspace}

\newcommand{\NumFilterLists}{8\xspace}
\newcommand{\CrawlerDwellSecs}{120\xspace}

\newcommand{\DataSetDomainsSuccesfullyCrawled}{8,998\xspace}

\newcommand{\DataSetNumImgResources}{201,785\xspace}
\newcommand{\DataSetNumImgResourcesAdFilterList}{11,584\xspace}
\newcommand{\DataSetNumImgResourcesAdAdGraph}{10,228\xspace}
\newcommand{\DataSetNumImgResourcesAccuracy}{98.95\%\xspace}
\newcommand{\DataSetNumImgResourcesPrecision}{93.09\%\xspace}
\newcommand{\DataSetNumImgResourcesRecall}{88.29\%\xspace}
\newcommand{\DataSetNumImgResourcesFPR}{0.39\%\xspace}
\newcommand{\DataSetNumImgResourcesFNR}{11.71\%\xspace}

\newcommand{\DataSetNumVideoResources}{2,360\xspace}
\newcommand{\DataSetNumVideoResourcesAdFilterList}{23\xspace}
\newcommand{\DataSetNumVideoResourcesAdAdGraph}{14\xspace}
\newcommand{\DataSetNumVideoResourcesAccuracy}{99.57\%\xspace}
\newcommand{\DataSetNumVideoResourcesPrecision}{93.33\%\xspace}
\newcommand{\DataSetNumVideoResourcesRecall}{60.86\%\xspace}
\newcommand{\DataSetNumVideoResourcesFPR}{0.04\%\xspace}
\newcommand{\DataSetNumVideoResourcesFNR}{39.14\%\xspace}

\newcommand{\DataSetNumScriptResources}{167,533\xspace}
\newcommand{\DataSetNumScriptResourcesAdFilterList}{67,959\xspace}
\newcommand{\DataSetNumScriptResourcesAdAdGraph}{60,030\xspace}
\newcommand{\DataSetNumScriptResourcesAccuracy}{90.52\%\xspace}
\newcommand{\DataSetNumScriptResourcesPrecision}{88.32\%\xspace}
\newcommand{\DataSetNumScriptResourcesRecall}{88.33\%\xspace}
\newcommand{\DataSetNumScriptResourcesFPR}{7.97\%\xspace}
\newcommand{\DataSetNumScriptResourcesFNR}{11.67\%\xspace}

\newcommand{\DataSetNumIFrameResources}{20,091\xspace}
\newcommand{\DataSetNumIFrameResourcesAdFilterList}{7,745\xspace}
\newcommand{\DataSetNumIFrameResourcesAdAdGraph}{7,244\xspace}
\newcommand{\DataSetNumIFrameResourcesAccuracy}{94.50\%\xspace}
\newcommand{\DataSetNumIFrameResourcesPrecision}{92.31\%\xspace}
\newcommand{\DataSetNumIFrameResourcesRecall}{93.53\%\xspace}
\newcommand{\DataSetNumIFrameResourcesFPR}{4.88\%\xspace}
\newcommand{\DataSetNumIFrameResourcesFNR}{6.47\%\xspace}

\newcommand{\DataSetNumAjaxResources}{24,365\xspace}
\newcommand{\DataSetNumAjaxResourcesAdFilterList}{8,305\xspace}
\newcommand{\DataSetNumAjaxResourcesAdAdGraph}{7,442\xspace}
\newcommand{\DataSetNumAjaxResourcesAccuracy}{93.55\%\xspace}
\newcommand{\DataSetNumAjaxResourcesPrecision}{91.31\%\xspace}
\newcommand{\DataSetNumAjaxResourcesRecall}{89.60\%\xspace}
\newcommand{\DataSetNumAjaxResourcesFPR}{4.40\%\xspace}
\newcommand{\DataSetNumAjaxResourcesFNR}{10.40\%\xspace}

\newcommand{\DataSetNumCSSResources}{124,207\xspace}
\newcommand{\DataSetNumCSSResourcesAdFilterList}{9,255\xspace}
\newcommand{\DataSetNumCSSResourcesAdAdGraph}{5,834\xspace}
\newcommand{\DataSetNumCSSResourcesAccuracy}{96.32\%\xspace}
\newcommand{\DataSetNumCSSResourcesPrecision}{83.61\%\xspace}
\newcommand{\DataSetNumCSSResourcesRecall}{63.03\%\xspace}
\newcommand{\DataSetNumCSSResourcesFPR}{0.99\%\xspace}
\newcommand{\DataSetNumCSSResourcesFNR}{36.97\%\xspace}

\newcommand{\DataSetNumResources}{540,341\xspace}
\newcommand{\DataSetNumResourcesAdFilterList}{104,871\xspace}
\newcommand{\DataSetNumResourcesAdAdGraph}{90,792\xspace}
\newcommand{\DataSetNumResourcesAccuracy}{95.33\%\xspace}

\newcommand{\AdGraphPrecision}{89.1\%\xspace}
\newcommand{\AdGraphRecall}{86.6\%\xspace}
\newcommand{\AdGraphAccuracy}{95.33\%\xspace}
\newcommand{\AdGraphFPR}{2.56\%\xspace}
\newcommand{\AdGraphFNR}{13.4\%\xspace}

\newcommand{\FalsePositiveNumSuccessImageCases}{282\xspace}
\newcommand{\FalsePositiveNumScriptCases}{100\xspace}

\newcommand{\FalsePositiveJSTruePositiveNum}{11\xspace}
\newcommand{\FalsePositiveJSTruePositivePct}{11.0\%\xspace}
\newcommand{\FalsePositiveJSFalsePositiveNum}{63\xspace}
\newcommand{\FalsePositiveJSFalsePositivePct}{63.0\%\xspace}
\newcommand{\FalsePositiveJSMixedNum}{22\xspace}
\newcommand{\FalsePositiveJSMixedPct}{22.0\%\xspace}
\newcommand{\FalsePositiveJSUndecideableNum}{4\xspace}
\newcommand{\FalsePositiveJSUndecideablePct}{4.0\%\xspace}

\newcommand{\FalsePositiveJSLowPct}{11.0\%\xspace}
\newcommand{\FalsePositiveJSHighPct}{33.0\%\xspace}
\newcommand{\FalsePositiveImgPct}{46.8\%\xspace}

\newcommand{\FalsePositiveImgTruePositiveNum}{132\xspace}
\newcommand{\FalsePositiveImgTruePositivePct}{46.8\%\xspace}
\newcommand{\FalsePositiveImgFalsePositiveNum}{129\xspace}
\newcommand{\FalsePositiveImgFalsePositivePct}{45.7\%\xspace}
\newcommand{\FalsePositiveImgSMixedNum}{0\xspace}
\newcommand{\FalsePositiveImgMixedPct}{0\%\xspace}
\newcommand{\FalsePositiveImgUndecideableNum}{21\xspace}
\newcommand{\FalsePositiveImgUndecideablePct}{7.4\%\xspace}

\newcommand{\FalseNegativeNumImageCases}{300\xspace}

\newcommand{\FalseNegativeNumScriptCases}{100\xspace}

\newcommand{\FalseNegativeJSTrueNegativeNum}{22\xspace}
\newcommand{\FalseNegativeJSTrueNegativePct}{22\%\xspace}
\newcommand{\FalseNegativeJSFalseNegativeNum}{55\xspace}
\newcommand{\FalseNegativeJSFalseNegativePct}{55\%\xspace}
\newcommand{\FalseNegativeJSMixedNum}{10\xspace}
\newcommand{\FalseNegativeJSMixedPct}{10\%\xspace}
\newcommand{\FalseNegativeJSUndecideableNum}{13\xspace}
\newcommand{\FalseNegativeJSUndecideablePct}{13\%\xspace}

\newcommand{\FalseNegativeJSLowPct}{22\%\xspace}
\newcommand{\FalseNegativeJSHighPct}{32\%\xspace}
\newcommand{\FalseNegativeImgPct}{27.7\%\xspace}

\newcommand{\FalseNegativeImgTrueNegativeNum}{83\xspace}
\newcommand{\FalseNegativeImgTrueNegativePct}{27.7\%\xspace}
\newcommand{\FalseNegativeImgFalseNegativeNum}{180\xspace}
\newcommand{\FalseNegativeImgFalseNegativePct}{60.0\%\xspace}
\newcommand{\FalseNegativeImgSMixedNum}{0\xspace}
\newcommand{\FalseNegativeImgMixedPct}{0\%\xspace}
\newcommand{\FalseNegativeImgUndecideableNum}{37\xspace}
\newcommand{\FalseNegativeImgUndecideablePct}{12.3\%\xspace}

\newcommand{\GroundTruthNumRulesEasyList}{72,660\xspace}
\newcommand{\GroundTruthNumRulesEasyPrivacy}{15,507\xspace}
\newcommand{\GroundTruthNumRulesAntiAdblock}{1,964\xspace}
\newcommand{\GroundTruthNumRulesWarningRemovalList}{378\xspace}
\newcommand{\GroundTruthNumRulesBlockzilla}{1,155\xspace}
\newcommand{\GroundTruthNumRulesFanboy}{38,675\xspace}
\newcommand{\GroundTruthNumRulesPeterLowe}{2,962\xspace}
\newcommand{\GroundTruthNumRulesSquid}{4,485\xspace}

\newcommand{\PerfAvgVsAdBlock}{78\%\xspace}
\newcommand{\PerfAvgVsChrome}{42\%\xspace}

\newcommand{\TotalNumOfDataSet}{110\xspace}
\newcommand{\AgreementRatio}{87.7\%\xspace}
\newcommand{\AGNoBreakageNum}{93.5\xspace}
\newcommand{\AGNoBreakagePct}{85.0\%\xspace}
\newcommand{\AGBreakagePct}{15.0\%\xspace}
\newcommand{\AGMajorBreakageNum}{6.5\xspace}
\newcommand{\AGMajorBreakagePct}{5.9\%\xspace}
\newcommand{\AGMinorBreakageNum}{7.5\xspace}
\newcommand{\AGMinorBreakagePct}{6.8\%\xspace}
\newcommand{\AGCrashBreakageNum}{2.5\xspace}
\newcommand{\AGCrashBreakagePct}{2.3\%\xspace}
\newcommand{\FLNoBreakageNum}{97.5\xspace}
\newcommand{\FLNoBreakagePct}{88.6\%\xspace}
\newcommand{\FLBreakagePct}{11.4\%\xspace}
\newcommand{\FLMajorBreakageNum}{7\xspace}
\newcommand{\FLMajorBreakagePct}{6.4\%\xspace}
\newcommand{\FLMinorBreakageNum}{4\xspace}
\newcommand{\FLMinorBreakagePct}{3.6\%\xspace}
\newcommand{\FLCrashBreakageNum}{1.5\xspace}
\newcommand{\FLCrashBreakagePct}{1.4\%\xspace}

\newcommand{\PrecisionDropWithoutStructuralFeatures}{6.6\%\xspace}
\newcommand{\RecallDropWithoutStructuralFeatures}{8.7\%\xspace}
\newcommand{\AccuracyDropWithoutStructuralFeatures}{2.7\%\xspace}

\newcommand{\AdGraphStkWalkPrecisonImprv}{1.5\%\xspace}
\newcommand{\AdGraphStkWalkRecallImprv}{16\%\xspace}
\newcommand{\AdGraphStkWalkAccuracyImprv}{2.3\%\xspace}

\newcommand{\AGStockIframeBetter}{37.66\%\xspace}
\newcommand{\AGStockScriptBetter}{20.82\%\xspace}
\newcommand{\AGStockImageBetter}{24.59\%\xspace}
\newcommand{\AGStockCSSBetter}{6.47\%\xspace}
\newcommand{\AGStockAJAXBetter}{48.03\%\xspace}
\newcommand{\AGStockVideoBetter}{7.14\%\xspace}

\newcommand{\AGStockIframeWorse}{30.47\%\xspace}
\newcommand{\AGStockScriptWorse}{17.96\%\xspace}
\newcommand{\AGStockImageWorse}{14.92\%\xspace}
\newcommand{\AGStockCSSWorse}{0.79\%\xspace}
\newcommand{\AGStockAJAXWorse}{36.14\%\xspace}
\newcommand{\AGStockVideoWorse}{6.20\%\xspace}

\begin{document}
 


\title{\Large \bf \tool: A Graph-Based Approach to Ad and Tracker Blocking}

\author{
{\rm Umar Iqbal$^*$$^\dag$ ~~ Peter Snyder$^\dag$ ~~ Shitong Zhu$^\ddag$ ~~ Benjamin Livshits$^\dag$$^\P$ ~~ Zhiyun Qian$^\ddag$ ~~ Zubair Shafiq$^*$}\\\\
$^*$University of Iowa ~~ $^\dag$Brave Software ~~ $^\ddag$UC Riverside ~~ $^\P$Imperial College London
} 

\maketitle
\thispagestyle{plain}
\pagestyle{plain}

\begin{abstract}

User demand for blocking advertising and tracking online is large and growing.
    Existing tools, both deployed and described in research, have proven
    useful, but lack either the completeness or robustness needed for a general
    solution.  Existing detection approaches generally focus on only one aspect
    of advertising or tracking (e.g. URL patterns, code structure), making
    existing approaches susceptible to evasion.

In this work we present \tool, a novel graph-based machine learning approach for detecting
    advertising and tracking resources on the web.  \tool differs from existing
    approaches by building a graph representation of the HTML structure,
    network requests, and \JS behavior of a webpage, and using this unique representation 
    to train a classifier for identifying advertising and tracking resources.  Because \tool considers many aspects of
    the context a network request takes place in, it is less susceptible to the
    single-factor evasion techniques that flummox existing approaches.

We evaluate \tool on the Alexa top-10K websites, and find that it is highly accurate,
    able to replicate the labels of human-generated filter lists with
    \DataSetNumResourcesAccuracy accuracy, and can even identify many mistakes in filter lists. We implement \tool as a modification
    to Chromium.  \tool adds only minor overhead to page loading and execution, and is actually faster than stock Chromium on
    \PerfAvgVsChrome of websites and AdBlock Plus on
    \PerfAvgVsAdBlock of websites.
    Overall, we conclude that \tool is both accurate enough and performant enough for online use, 
    breaking comparable or fewer websites than popular filter list based approaches.
\end{abstract}

\section{Introduction}

The need for content blocking on the web is large and growing. Prior
research has shown that blocking advertising and tracking resources improves performance~\cite{Garimella17WebSciPerformance,Pujol15AnnoyedUsersIMC,NYTimesAdblockPerformance}, privacy \cite{Lerner16InternetJonesUSENIX,Englehardt16MillionSiteMeasurementCCS,anthes15databrokerscacm}, and
security~\cite{Merzdovnik17BlockMeIfYouCanESP,Gervais17ESORICSQuantifyingPrivacy}, in addition to making the browsing experience more
pleasant~\cite{pagefair_report}. 
Browser
vendors are increasingly integrating content blocking into their browsers \cite{safari_itp,mozilla_tracker_blocking,chrome_adblock},
and user demand for content blocking is expected to grow in future~\cite{pagefair_desktop,pagefair_mobile}.

While existing content blocking tools are useful, they are vulnerable to
practical, realistic countermeasures. Current techniques generally block
unwanted content based on URL patterns (using manually-curated filter lists which contain rules that
describe suspect URLs), or patterns in \JS behavior or code structure. Such
approaches fail against adversaries who rotate domains
quickly \cite{dga_blog}, proxy resources through trusted domains (e.g. the first party, CDNs) \cite{instartlogic},
or restructure or obfuscate \JS \cite{LeMN17AccurateDetectionObfusctaion}, among other common techniques.

As a result, researchers have proposed several alternative approaches
to content blocking. While these approaches are interesting, they
are either incomplete or susceptible to trivial circumvention from
even mildly determined attackers. Existing proposals suggest filter lists, pre-defined heuristics, 
and machine learning (ML) approaches that leverage network or code analysis for identifying unwanted web
content, but fail to consider enough context to avoid trivial evasions.

This work presents \tool, an accurate and performant graph-based ML approach for detecting and blocking unwanted (advertising and tracking) resources on the web.
\tool makes blocking decisions using a novel graph representation of a webpage's past and present HTML structure, the behavior and interrelationships
of all executed \JS code units, and the destination and cause of all network
requests that have occurred up until the considered network request.
This contextually-rich blocking approach allows \tool to both identify
unwanted resources that existing approaches miss, and makes \tool more
robust against simple evasions that flummox existing approaches.

\tool is designed for both online (i.e. in-browser, during page execution) and 
offline (i.e. for filter list construction) deployment. 
\tool is performant enough for online deployment; its performance is comparable to stock Chromium and better than Adblock Plus.
\tool can also be used offline to create or augment filter lists used by extension-based content blocking approaches.
This dual deployment strategy can benefit users of \tool directly as well as users of extension-based content blocking approaches.

This work makes the following contributions to the problem of identifying
and blocking advertising and tracking resources on the web.

\begin{enumerate}

\item A \textbf{graph-based ML approach} to identify advertising and
        tracking resources in websites based on the HTML structure, \JS behavior,
        and network requests made during execution.

\item A \textbf{large scale evaluation} of \tool's ability to detect advertising
        and tracking resources on popular websites.  We find that
        \tool is able to replicate the labels of human-generated filter
        lists with \AdGraphAccuracy accuracy.  Further, \tool is able to
        outperform existing filter lists in many cases, by correctly
        distinguishing ad/tracker resources from benign resources in cases
        where existing filter lists err.

\item A \textbf{performant implementation} of \tool as a
        patch to Chromium.\footnote{ Since \tool is 
        designed and implemented in Chromium, it can be readily deployed
         on other Chromium based browsers (e.g. Chrome, Brave).} Our approach modifies the Blink and V8 components
        in Chromium to instrument and attribute document behavior in a way
        that exceeds existing practical approaches, without significantly
        affecting browser performance. \tool loads pages faster than
        stock Chromium on \PerfAvgVsChrome of pages, and faster than
        AdBlock Plus on \PerfAvgVsAdBlock of pages.

\item A \textbf{breakage analysis} of \tool's impact on popular websites.
        \tool has a noticeable negative affect on benign page
        functionality at rates similar to filter lists
        (affecting \AGBreakagePct versus \FLBreakagePct of websites
        respectively) and majorly affects page functionality less than 
        filter lists (breaking \AGMajorBreakagePct versus \FLMajorBreakagePct
        websites, respectively).

\end{enumerate}

The rest of this paper is structured as follows.  Section~\ref{sec:background}
presents existing work on the problem of ad and tracker blocking, and discusses why existing approaches are insufficient as comprehensive blocking
solutions.  Section~\ref{sec:design} describes the design and implementation
of \tool. Section~\ref{sec:eval} presents an evaluation of \tool's
effectiveness as a content blocking solution, in terms of blocking
accuracy, performance, and effect on existing websites.
Section~\ref{sec:discussion} describes \tool's limitations, how \tool can be
further improved, and potential uses for \tool in offline scenarios.  Section~\ref{sec:conclusion}
concludes the paper.

\section{Background and Related Work}
\label{sec:background}

\subsection{Problem Difficulty}
\label{sec:background:difficulty}

Ad and tracker blocking is a well studied topic (e.g. \cite{Bau13PromisingTrackerW2SP,Yu16TrackingTheTrackersWWW,Wu16MLTrackingESORICS,Shuba2018PETSNoMoAds,Kaizer16JSTrackingIWSPA,Ikram17SeamlessTrackingPETS,Gugelmann15ComplementBlacklistPETS,Bhagavatula14MLforFilterListsAISec}).  
However, existing work is insufficient to form a comprehensive and robust blocking solution.

Many existing approaches (e.g. \cite{Gugelmann15ComplementBlacklistPETS,Bhagavatula14MLforFilterListsAISec}) are vulnerable to commonly deployed countermeasures,
such as evading domain-based blocking through domain generation algorithms (DGA) \cite{dga_blog},
hosting tracking related code on the first-party domain \cite{instartlogic}, spreading tracking
related behavior across multiple code units, and code obfuscation \cite{LeMN17AccurateDetectionObfusctaion}. Much related work in the area is unable to reason about domains that
host both ``malicious'' (ads and tracking) and ``benign'' (functional or user desireable) content, and end up over or under labeling
resources.

Other existing work (e.g. \cite{Bau13PromisingTrackerW2SP,Kaizer16JSTrackingIWSPA}) lacks realistic evaluations. Sometimes
this takes the form of an ambiguous comparison to ground truth 
(making it challenging to ascertain the usefulness of the technique as a deployable solution).
Other cases target advertising or tracking, but not both
together.  Still other cases target only a subset of advertising or tracking
related resources (e.g. scripts or images), but fail to consider other
ways advertising or tracking can be carried out (e.g. iframes and CSS styling 
rules).

Further existing  work (e.g. \cite{Wu16MLTrackingESORICS,Ikram17SeamlessTrackingPETS}) presents a strategy for blocking resources, but lacks an
evaluation of how much benign (i.e. user desirable) functionality the approach
would break. This leaves a proposal for preventing a subset of an application's
code from executing, without an understanding of how it effects the functioning
of the overall application (user-serving or otherwise).  These approaches
may fail to separate the wheat from the chaff; they may prevent advertising
and tracking, but at the expense of breaking desirable functionality.

The rest of this section reviews existing work
on blocking advertising and tracking content on the web.  Emphasis is given
both on the contributions of each work, and why each work is incomplete
as a deployable, real-world blocking solution.

\subsection{Existing Blocking Techniques}
\label{sec:background:blocking}
\begin{table*}[htbp]
    \begin{center}
        \begin{tabular}{r|c|c|c|c|c|c|c} 
            \toprule
                Approach     & Ad/Tracker & Domain/URL & 1st,3rd Party & DGA            & Code Structure  & Cross JS Collaboration  & Breakage \\ 
                             & Blocking  & Blocking   & Blocking          & Susceptibility & Susceptibility  &  Susceptibility         & Analysis \\ 
            \midrule
                Bau et al. \cite{Bau13PromisingTrackerW2SP}          & Tracker     & Domain & 3rd party            & Yes & -   & -    & No (-) \\ 
                Yu et al. \cite{Yu16TrackingTheTrackersWWW}         & Tracker     & Domain & 3rd party             & Yes & No  & No   & Yes (25\%) \\ 
                Wu et al. \cite{Wu16MLTrackingESORICS}              & Tracker     & Domain & 3rd party            & Yes & No  & Yes  & No (-) \\
                Shuba et al. \cite{Shuba2018PETSNoMoAds}               & Ads         & URL    & 1st,3rd party & Yes & No  & No   & No (-) \\
                Kaizer and Gupta \cite{Kaizer16JSTrackingIWSPA}            & Tracker     & Domain & 3rd party            & Yes & Yes & Yes  & No (-) \\ 
                Ikram et al. \cite{Ikram17SeamlessTrackingPETS}        & Tracker     & URL    & 1st,3rd party & No  & Yes & Yes  & No (-) \\ 
                Gugelmann et al. \cite{Gugelmann15ComplementBlacklistPETS} & Ads,Tracker & Domain & 3rd party             & Yes & No  & Yes  & No (-) \\  
                Bhagavatula et al. \cite{Bhagavatula14MLforFilterListsAISec} & Ads         & URL    & 1st,3rd party & Yes & No  & No   & No (-) \\ 
            \bottomrule
            \multicolumn{8}{c}{}
        \end{tabular}
        \caption{Comparison of the related work, including the practical evasions and countermeasures each is vulnerable to. Ad/Tracker Blocking column represents blocking of ads, trackers, or both. Domain/URL Detection column represents blocking at domain or URL level. 1st,3rd Party Blocking column, represents blocking of third-party requests, first-party requests, or both. In DGA Susceptibility, Code Structure Susceptibility, and Cross JS Collaboration Susceptibility columns, Yes and No represent that the approach's susceptibility to specified countermeasure. The Breakage Analysis column represents whether the breakage analysis was performed by the approach and their results.}
        \label{table: related work}
        \vspace{-10pt}
    \end{center}
\end{table*}    

This subsection describes existing tracking and advertising blocking work,
categorized by the types of evasions each approach is vulnerable to.
Our goal is not to lessen the contributions of existing work (which
are many and significant), but merely to highlight the kinds of practical
and deployed evasions each is vulnerable to, to further motivate the need
for a more comprehensive solution.

Note that many blocking approaches discussed here are vulnerable to multiple evasions.  
In these cases, we discuss only one category of evasion the work is vulnerable to.  
Table~\ref{table: related work} summarily compares the strengths and weaknesses of existing approaches.

\point{Domain Based Blocking}
Many existing content blocking approaches attempt to prevent advertising and tracking by
identifying suspect domains (eTLD+1), and blocking all requests to
resources on such domains. These approaches are insufficient for several
reasons.  First, determined advertising and tracking services can use DGA
to serve their content from quickly changing domains that are
unpredictable to the client, but known to the adversary.  Such evasions
trivially circumvent approaches that depend primarily, or only, on
domain blocking strategies \cite{dga_blog}. Similarly, in many cases, domain-focused
approaches are easily circumvented by proxying the malicious resource through
the first-party domain \cite{instartlogic}. A comprehensive blocking solution should be able to
account for both of these evasion strategies.

AdBlock Plus~\cite{adblockplus_web}, uBlock Origin~\cite{ublockOrigin_web}, Ghostery~\cite{ghostery_web},
and Disconnect~\cite{disconnect_me} are all popular and deployed solutions that depend
solely or partially on the domain of the request, and are thus vulnerable to
the above discussed approaches.  These approaches use filter lists, which
describe hosts, paths, or both of advertising and tracking resources.

Gugelmann et al.~\cite{Gugelmann15ComplementBlacklistPETS} developed a ML-based approach for augmenting filter lists, by using existing filter lists
as ground truth, and training a classifier based on the HTTP and domain-request
behavior of additional network requests.  Bhagavatula et
al.~\cite{Bhagavatula14MLforFilterListsAISec} developed a ML-based approach for generating future domain-and-path based filter lists,
using the rules in existing filter lists as ground truth.  These approaches
may be useful in identifying additional suspect content, but are easily
circumvented by an attacker willing to take any of the domain hiding or rotating
measures discussed earlier.

Yu et al.~\cite{Yu16TrackingTheTrackersWWW} described a method for detecting
tracking related domains by looking for third-parties that receive similar
unique tokens across a significant number of first-parties.  This approach
hinges on an attacker using the same receiving domain over a large number of
hosting domains.  Apple's Safari browser includes a similar technique called
Intelligent Tracking Protection~\cite{safari_itp}, that identifies tracking
related domains by looking for third-party contexts that access state without
user interaction. Privacy Badger~\cite{privacybadger_web} also identifies tracking
related domains by looking for third-party domains that track users (e.g., by setting identifying 
cookies) on three or more sites. These techniques do not attempt to block advertising, and
also require that the attacker use consistent domains.  Bau et
al.~\cite{Bau13PromisingTrackerW2SP} proposed building a graph of
resource-hosting domains and training a ML classifier based on commonalities of
third-party hosted code, again relying on hosting domains being distinct,
consistent, and long lasting.

\point{\JS Code Unit Classification}
Other blocking approaches attempt to identify undesirable code based on the
structure or behavior of \JS code units.  Such approaches take as input a single
code unit (and sometimes the resulting behavior of that code unit), and train ML classifiers for identifying undesirable code.

Blocking approaches that rely solely on \JS behavior or structure are vulnerable
to several easy to deploy countermeasures.  Most trivially, these approaches do
not consider the interaction between code units.  An attacker can easily
avoid detection by spreading the malicious behavior across multiple code units,
having each code unit execute a small enough amount of suspicious behavior to
avoid being classified as malicious, and then using a final code unit to combine
the quasi-identifiers into a single exfiltrated value.  Examples of such work
includes the approaches given by Wu et al.~\cite{Wu16MLTrackingESORICS} and
Kaiser et al.~\cite{Kaizer16JSTrackingIWSPA}, both of whom propose ML
classifiers that take as input the DOM properties accessed by \JS (among other
things) to determined whether a code unit is tracking related.

Other approaches attempt to identify tracking-related \JS based on the
static features of the code, such as names of cookie values, or similar
sub-sections in the code.  Such approaches are vulnerable to many obfuscation
techniques, including using \JS's dynamic nature to break identifying strings
and labels up across a code base, using dynamic interpretation facilities
in the language (e.g. \texttt{eval}, \texttt{new Function}) to confuse
static detection, or simply using different parameters for popular \JS
post-processing tools (e.g. JSMin \cite{JSMin},
Browserify \cite{Browserify}, Webpack \cite{Webpack},
RequireJS \cite{RequireJS}). Ikram et
al.~\cite{Ikram17SeamlessTrackingPETS} proposed one such vulnerable technique, by training
a ML classifier to identify static features in \JS code labeled by existing
filters lists as being tracking related, and using the resulting model
to predict whether future \JS code is malicious.

\point{Evaluation Issues}
Much related work lacks a comprehensive and realistic evaluation.  
Examples include ambiguous or unstated sources of ground truth comparison (e.g.
\cite{Bau13PromisingTrackerW2SP}), unrealistic metrics for what constitutes tracking or non-tracking \JS
code (e.g. \cite{Ikram17SeamlessTrackingPETS} makes the odd assumption
that \JS code that tracks mouse or keyboard behavior is automatically benign,
despite the most popular tracking libraries including the ability to track such
functionality \cite{GoogleAnalytics}),
or the decision to (implicitly or explicitly) whitelist all first-party
resources (e.g. \cite{Bau13PromisingTrackerW2SP}, \cite{Yu16TrackingTheTrackersWWW}, \cite{Wu16MLTrackingESORICS}, \cite{Kaizer16JSTrackingIWSPA},
\cite{Gugelmann15ComplementBlacklistPETS}).

More significantly, much related work proposes resource blocking strategies,
but without an evaluation of how their blocking strategy would affect the
usability of the web. To name some examples, \cite{Bau13PromisingTrackerW2SP},
\cite{Wu16MLTrackingESORICS}, \cite{Shuba2018PETSNoMoAds},
\cite{Kaizer16JSTrackingIWSPA}, \cite{Ikram17SeamlessTrackingPETS},
\cite{Gugelmann15ComplementBlacklistPETS}, and
\cite{Bhagavatula14MLforFilterListsAISec}, all propose
strategies for automatically blocking web resources in pages, without
determining whether that blocking would harm or break the user-serving goals of
websites (\cite{Yu16TrackingTheTrackersWWW} is an laudable exception,
presenting an indirect measure of site breakage by way of how often users disabled their tool when browsing). 
Work that presents how much \textit{bad} website behavior an approach avoids, without also
presenting how much \textit{beneficial} behavior the approach breaks, is
ignoring one half of the ledger, making it difficult to evaluate each work
as a practical, deployable solution.

\subsection{\JS Attribution}
\label{sec:background:attribution}
We next present existing work on a related problem of
attributing DOM modifications to responsible \JS code units.  \JS attribution is a
necessary part of the broader problem of blocking ads and trackers, 
as its necessary to trace DOM modifications and network requests 
back to their originating \JS code units.  Without attribution, it is difficult-to-impossible
to understand which party (or element) is responsible for which undesired activity.

While there have been several efforts to build systems to attribute DOM modifications to \JS code
units, both in peer-reviewed literature and in deployed software, all
existing approaches suffer from completeness and correctness issues. 
Below we present existing \JS attribution approaches and discuss why they are lacking.

 
\point{\JS Stack Walking}
The most common \JS attribution technique is to interpose on the prototype chain
of the methods being observed, throw an exception, and walk the resulting stack
object to determine what code unit called the modified (i.e. interposed on)
method. This technique is used, for example, by Privacy Badger \cite{privacybadger_web}. The technique has
the benefit of not requiring any browser modifications, and of being able to run
``online'' (e.g. the attribution information is available during execution, allowing
for runtime policy decisions).

Unfortunately, stack walking suffers from correctness and completeness issues.
First, there are many cases where calling code can mask its identity from the
stack, making attribution impossible. Examples include eval'ed code and
functions the \JS runtime decides to inline for performance purposes. Malicious
code can be structured to take advantage of these shortcomings to evade
detection \cite{Curtsinger11ZozzleUSENIX}.

Second, stack walking requires that code be able to modify the prototype objects
in the environment, which further requires that the attributing (stack walking)
code run before any other code on the page. If untrusted code can gain
references to unmodified data structures (e.g. those not interposed on by the
attributing code), then the untrusted code can again avoid detection. Browsers
do not currently provide any fool-proof way of allowing trusted code to restrict
untrusted code from accessing unmodified \DOM structures. For example, untrusted
code can gain access to unmodified \DOM structures by injecting subdocuments and
extracting references to from the subdocument, before the attributing code can
run in the subdocument.

\point{AdTracker}
Recent versions of Chromium include a \JS attribution system called
AdTracker \cite{ChromeAdTracker},
which attributes \DOM modifications made in the Blink rendering system to \JS code
execution in \VJS, the browser's \JS engine. AdTracker is used by Chromium to
detect when \TP code modifies the \DOM in a way that violates Google's ad
policy \cite{chrome_adblock}, such as when \JS code creates large overlay elements across the page.
The code allows the browser to determine which code unit on the page is
responsible for the violating changes, instead of holding the hosting page
responsible.

AdTracker achieves correctness but lacks completeness. In other words, the
cases where AdTracker can correctly do attribution are well defined, but there
are certain scenarios where AdTracker is not able to maintain attribution. 
At a high level, AdTracker can do attribution in \textit{macrotasks}, but not in \textit{microtasks}. 
Macrotasks are a subset of cases where \VJS is invoked by Blink or when one
function invokes another within \VJS. Microtasks can be thought of as an
inlining optimization used by \VJS to save stack frames, and is used in cases
like callback functions in native \JS APIs (e.g. callback functions to
\texttt{Promises}). Effectively, AdTracker trades completeness for
performance,\footnote{These shortcomings are known to the
Chromium developers, and are an intentional tradeoff to maximize
performance.} which means that a trivial code transformation 
can circumvent AdTracker. 

\point{JSGraph}
JSGraph \cite{LiNDSS18JSgraph} is designed for offline \JS attribution. At a
high level, JSGraph instruments locations where control is exchanged between
Blink and \VJS, noting which script unit contains the function being called, and
treating all subsequent \JS functionality as resulting from that script unit. At
the next point of transfer from Blink to \VJS, a new script unit is identified,
and following changes are attributed to the new script.

JSGraph writes to a log file, which makes it potentially useful for certain
types of offline forensic analysis, but not useful for online content blocking.
More significantly, JSGraph suffers from correctness and completeness issues. First, like AdTracker,
JSGraph does not provide attribution for functionality optimized into
microtasks. Second, JSGraph's attribution provides incorrect results 
(e.g. unable to link eval'ed created script in a callback to its parent script) 
in the face of other \VJS optimizations, such as deferred parsing, where \VJS compiles
different sections of a single script unit at different times. Third, JSGraph
mixes all frames and subframes loaded in a page together, causing confusion as
to which script is making which changes (the script unit identifier used by
JSGraph is re-used between frames, so different scripts in different frames can
have the same identifier in the same log file).

\section{\tool Design}
\label{sec:design}

\begin{figure*}[!htp]
      \centering
      \includegraphics[width=\textwidth]{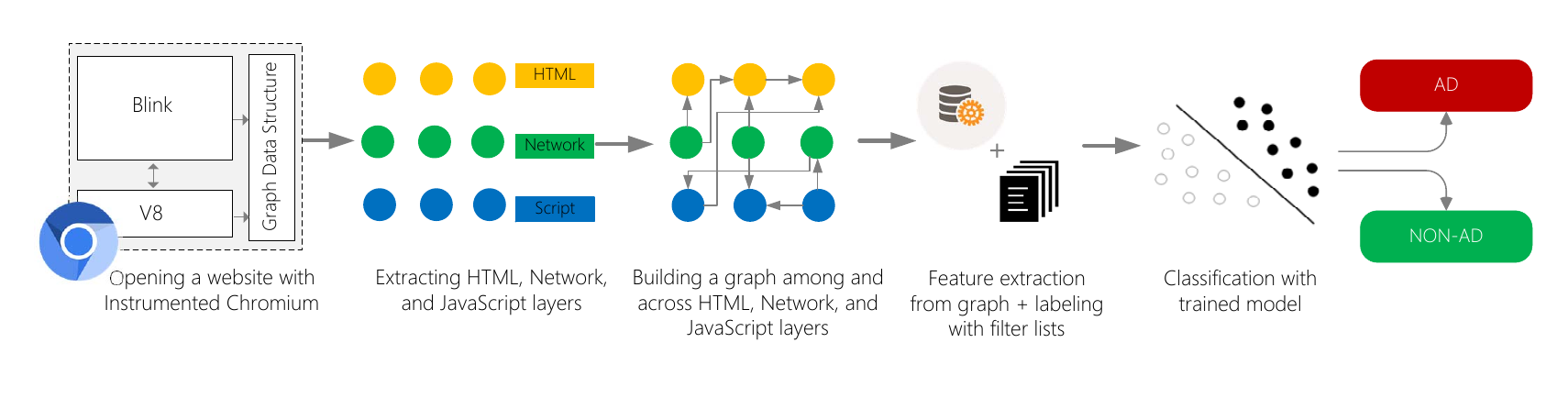}
      \caption{\tool: Our proposed approach for ad and tracking blocking. We instrument Chromium to extract 
      information from HTML structure, network, and JavaScript behavior of a webpage execution as a graph representation. 
      We then extract distinguishing structural and content features from the graph and train a ML model to detect ads and trackers.}
      \label{figure:classifier_flow}
      \vspace{-10pt}
\end{figure*}

In this section we present the design and implementation of \tool, an in-browser
ML-based approach to block ad and tracking related content on the web. 
We first describe a novel graph representation of the execution
of a website that tracks changes in the HTML structure, behavior and
interaction between \JS code, and network requests of the page over time.  This
graph representation allows for tracing the provenance of any DOM change to the 
responsible party (e.g. \JS code, the parser, a network request). 
Second, we discuss the Chromium instrumentation needed to construct our graph
representation. Third, we describe the features \tool extracts from our graph
representation to distinguish between ad/tracker and benign resources. Finally, we explain
the supervised ML classifier and how \tool enforces its
classification decisions at runtime. Figure \ref{figure:classifier_flow} gives
an architectural overview of \tool.

\subsection{Graph Representation}
\label{sec:classifier:representation}

Webpages are parsed and represented as DOM trees in modern browsers.
The DOM tree captures relationships among HTML elements (e.g. parent-child, sibling-sibling).
In \tool, we enrich this existing tree-representation with additional information about the 
execution and communication of the page, such as edges to capture JavaScript's interactions with
HTML elements, or which code unit triggered a given network request.
These edge additions transform the DOM tree to a graph.
\tool uses this graph representation to capture the execution of a webpage.

\tools graph representation of page execution tracks changes 
in the website's HTML structure, network requests, and \JS behavior.
The unique graph structure brings several benefits.  First, because the
graph contains information about the cause and content of every network
request and DOM modification during the page's life cycle, the graph
allows for tracing the provenance of any change or behavior back to either
the responsible \JS code unit, or, in the case of initial HTML text, the browser's
HTML parser.  Second, the graph representation allows for 
extraction of context-rich features, which are used by \tool to identify advertising
and tracking related network requests. For example, the graph allows for quick determinations of the
source script sending an \AJAX request, the position, depth, and location of an
image request, and whether a subdocument was injected in a page from \JS code,
among many others. The contextual information captured by these features in \tool far
exceeds what is available to existing blocking tools, as discussed in Section \ref{sec:background}.

Next, we explain how \tool represents information during a page load as nodes
and edges in a graph.

\point{Nodes}
\tool depicts all elements in a website as one of four types of node:
\textit{parser}, \textit{HTML}, \textit{network}, or \textit{script}.

The parser node is a single, special case node that \tool uses to
attribute document changes and network requests to the HTML parser, instead of
script execution. Each graph contains exactly one parser node.

HTML nodes represent HTML elements in the page, and map directly onto the
kinds of tags and markup that exist in websites. Examples of HTML nodes
include image tags, anchor tags, and paragraph tags.
HTML nodes are annotated to store information about the tag type and the
tags HTML attributes (e.g. \texttt{src} for image tags, \texttt{class} and
\texttt{id} for all tags, and \texttt{value} for input tags). HTML text nodes
are represented as a special case HTML node, one without a tag
type.

Network nodes represent remote resources, and are annotated with the
type of resource being requested. Requests for sub-documents (i.e.
\texttt{iframes}), images, \texttt{XMLHTTPRequest} fetches, and
others are captured by network nodes.

Script nodes represent each compiled and executed body of \JS code in the
document. In most cases, these can be thought of as a special type of HTML
node, since most scripts in the page are tied to script tags (whether inline or
remotely fetched). \tool represents script as its own node type though to also
capture the other sources of script execution in a page (e.g.
\texttt{javascript:} URIs).

\point{Edges}
\tool uses edges to represent the relationship between any two nodes in the
graph. All edges in \tool are directed. Depending on the execution of pages,
the graph may contain cycles. All edges in \tool are of one of three types,
\textit{structural}, \textit{modification}, and \textit{network}.

Structural edges describe the relationship between two HTML elements on a page
(e.g. two HTML nodes). Mirroring the \DOM API, edges are inserted to
describe parent-child node relationships, and the order of sibling nodes.

Modification edges depict the creation, insertion, removal, deletion, and
attribute modification of each HTML node. Each modification
edge notes the type of event (e.g. node creation, node modification, etc) and
any additional information about the event (e.g. the attributes that were modified, their
new values, etc). Each modification edge leaves a script or parser node, and
points to the HTML element being modified.

Network edges depict the browser making a request for a remote resource
(captured in the graph as a network node). Network edges leave
the script or HTML node responsible for the request being made, and point
to the network node being requested. Network edges are
annotated with the URL being requested.

\point{Composition Examples}
These four node types and three edge types together depict changes to DOM state in
a website. For example, \tool represents an HTML tag \texttt{<img
src="/example.png">} as an HTML node depicting the \texttt{img} tag, a
network node depicting the image, and a network edge, leaving
the former and pointing to the latter, annotated with the
``\texttt{/example.png}'' URL.  
As another example, a script modifying the value of a form element would be represented as
a script node depicting the relevant \JS code, an HTML node describing the
form element being modified, and a modification edge describing a modification
event, and the new value for the ``value'' attribute.

\subsection{Graph Construction}
\label{sec:classifier:construction}
\tool's graph representation of page execution requires low level modifications
to the browser's fetching, parsing, and \JS layers. We implement \tool as a
modification to the Chromium web browser.\footnote{The source code of our Chromium implementation is available at: \\ \url{https://uiowa-irl.github.io/AdGraph/}.}
The Chromium browser consists of many
sub-projects, or modules. The Blink~\cite{BlinkRenderer} module is responsible
for performing network requests, parsing HTML, responding to most kinds of user
events, and rendering pages. The \VJS~\cite{ChromiumV8} module is responsible
for parsing and executing \JS. Next, we provide a high level overview of the
types and scope of our modifications in Chromium for constructing \tool's graph
representation.

\point{Blink Instrumentation}
We instrument Blink to capture anytime a network request is about to be sent, anytime a new HTML node is being created, deleted or otherwise
modified (and noting whether the change was due to the parser or \JS execution),
and anytime control was about to be passed to \VJS. We further modify each
page's execution environment to bind the graph representation of the page to
each page's document object. This choice allows us to easily distinguish
scripts executing in different frames/sub-documents, a problem that has
frustrated prior work (see discussion of JSGraph in Section~\ref{sec:background:attribution}).
Finally, we add instrumentation to allow us to map between \VJS's identifers
for script units, and the sources of script in the executing site (e.g. script
tags, eval'ed scripts, script executed by extensions).\footnote{The
architectural independence between the \VJS and Blink projects made this an
unexpectedly difficult problem to solve, with many unanticipated corner cases
that were not discovered until we subjected \tool to extensive automatic and
manual testing.}

\point{\VJS Instrumentation}
We also modify \VJS to add instrumentation points to allow us to track anytime a
script is compiled, and anytime control changes between script units. We
accomplish this by associating every function and global scope to the script
they are compiled from. We then can note every time a new scope is entered, and
attribute any document modifications or network requests to that script, until
the scope is exited.

\VJS contains several optimizations that make this general
approach insufficient. First, \VJS sometimes defers parsing of subsections of \JS code. A partial list
of such cases includes eval'ed code, code compiled with the \texttt{Function}
constructor, and anonymous functions provided as callbacks for some built in
functions (e.g. \texttt{setTimeout}). To handle these cases, \tool not only maps
functions to script units but also sub-scripts to scripts.

Second, \VJS implements microtasks that make attribution difficult.
Microtasks allow for some memory savings (much of the type information and
vtable look-up overhead is skipped) and reduce some book-keeping overhead.
Tracking attribution of DOM changes in microtasks is difficult because, at this
level, \VJS no longer tracks functions as C++ objects, but as compiled
bytecode, requiring a different approach to determining which script unit
``owned'' any given execution. \tool solves this problem through additional
instrumentation, and some runtime stack scanning, yielding completeness at the
cost of a minor performance overhead.

\point{\JS Attribution Example}
\tool is able to attribute DOM modifications and network events to
script units in cases where existing techniques fail.  We give a
representative example in code snippet \ref{lst:sample_html}. 

This code uses \texttt{eval} to parse and execute a string as \JS code. The
resulting code uses a \texttt{Promise} in a \texttt{setTimeout} callback.
This \texttt{Promise} callback is optimized in \VJS as a microtask, which evades
the attribution techniques used in current work (e.g. PrivacyBadger /
stack walking, AdTracker, JSGraph, discussed in Section \ref{sec:background:attribution}).
Existing tools would not be able to recognize that this code 
unit was responsible for the image fetched in the \texttt{Promise} callback.

\begin{figure}[htbp]
      \centering
      \includegraphics[width=.5\textwidth]{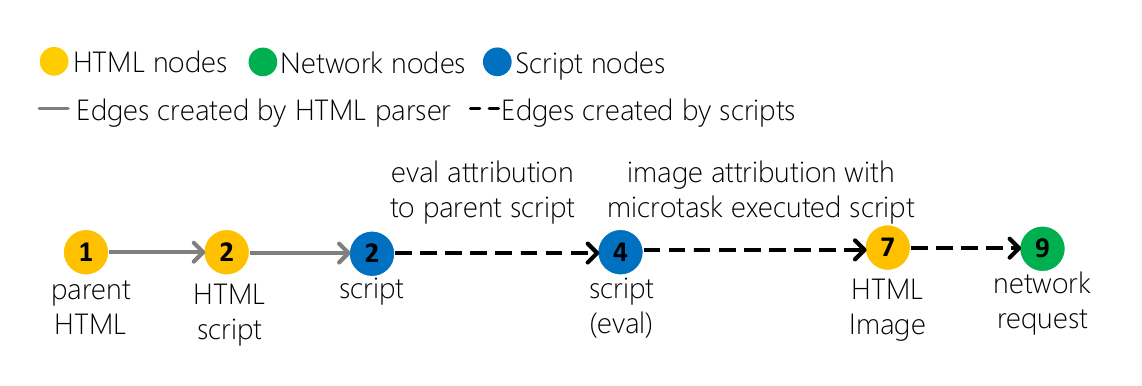}
      \caption{\tool's representation of example code snippet \ref{lst:sample_html}. Node numbers correspond to line numbers in code
      snippet \ref{lst:sample_html}.  This example highlights connections
      and attributions not possible in existing techniques.}
      \label{figure:example_sub_graph}
      \vspace{-10pt}
  \end{figure}

\tool, though, is able to correctly attribute the image request to this code
unit. Figure \ref{figure:example_sub_graph} shows how this execution pattern
would be stored in \tool.  Specifically, the edge between nodes 2 and 4 records
the attribution of the \texttt{eval} call to the responsible \JS code unit,
and the edge between nodes 7 and 9 in record that the image request is a result
of code executed in the microtask.  Existing approaches would either miss
the edge between 2 and 4, or 7 or 9.

\begin{figure}[!htpb]
\begin{lstlisting}[style=htmlcssjs,caption=A microtask in an eval created script loading an ad., label={lst:sample_html}]
<html>
   <script>
       ...
       eval("setTimeout(function xyz() {
       const p = Promise.resolve('A');
       p.then(function abc(_) {
         var img = document.createElement('img');
         img.setAttribute('id','ad_image');
         img.src = 'adnetwork.com/ad.png';
       }) }, 5) ");
       ...
   </script>
</html>
\end{lstlisting}
\end{figure}
\vspace{-10pt}

\subsection{Feature Extraction}
Next, we present the features that \tool extracts from the graph
to distinguish ads and trackers from functional resources.
These features are designed based on our domain knowledge and expert intuition.
%
%
Specifically, we manually analyze a large number of websites and try to
design features that would distinguish ad/tracking related
resources from functional (or benign) resources.

The extracted features broadly fall into two categories: ``structural''
(features that consider the relationship between nodes and edges in the
graph) and ``content'' (features that depend on the values and 
attributes of nodes in isolation from their connections). 
In total we extract \NumFeatures structural and content features.
Table \ref{table:features_list} gives a summary and representative examples of
features from each category.
Below we provide a high-level description of structural and content features. 
More detailed analysis of features and their robustness is presented in Section \ref{sec:eval:ffeatureanalysis}.

\begin{table}[h]
\centering
\begin{tabular}{lr}
    \toprule
        Structural Features \\
    \midrule
        Graph size (\# of nodes, \# of edges, and nodes/edge ratio) \\ 
        Degree (in, out, in+out, and average degree connectivity) \\
        Number of siblings (node and parents) \\
        Modifications by scripts (node and parents) \\
        Parent's attributes \\
        Parent degree (in, out, in+out, and average degree connectivity) \\
        Sibling's attributes \\
        Ascendant's attributes \\
        Descendant of a script \\
        Ascendant's script properties \\
        Parent is an eval script \\
    \midrule
        Content Features \\
    \midrule
        Request type (e.g. \texttt{iframe}, \texttt{image}) \\
        Ad keywords in request (e.g. banner, sponsor) \\
        Ad or screen dimensions in URL \\
        Valid query string parameters \\
        Length of URL \\
        Domain party \\
        Sub-domain check \\
        Base domain in query string \\ 
        Semi-colon in query string \\
    \bottomrule
    \\
\end{tabular}
\caption{Summarized feature set used by \tool.}
\label{table:features_list}
\vspace{-10pt}
\end{table}

\point{Structural Features}
Structural features target the relationship between elements in a page
(e.g. the relationship between a network request and the responsible script
unit, or a HTML nodes' parents, siblings and cousin HTML nodes). 
Examples of structural features include whether a node's \textit{parents} have
ad-related values for the \texttt{class} attribute, the tag names of the
node's \textit{siblings}, or how deeply nested in the document's structure
a given node is.

Structural features also consider the interaction between \JS code, and the
resource being requested. These features rely on \tool's instrumentation of Blink and \VJS. 
Examples of \JS features include whether the node initiating a network request was inserted by \JS code, the number of scripts that have ``touched'' the node issuing the request, and, in the case of requests that are not directly related to HTML elements (e.g. AJAX), whether the \JS code initiating the request was inlined in the document or fetched from a third-party.

\point{Content Features}
Content features relate to values attached to individual nodes in the graph
(and not the connections \textit{between} nodes in the graph).  The most 
significant value considered is the URL of the resource being requested. These
content features are similar to what most existing content blocking tools use.
\tool's specific set of features though is unique. Examples of \tool's content
features include whether the origin of the resource being requested is
first-or-third party, the number of path segments in the URL being requested,
and whether the URL contains any ad-related keywords.

\subsection{Classification}
\tool uses random forest \cite{Breiman01RandomForest}, a well-known ensemble supervised ML classification algorithm. 
Random forest combines decisions from multiple decision trees, each constructed using a different bootstrap sample of the data, by choosing the mode of the predicted class distribution.
Each node for a decision tree is split using the best among the subset of features selected at random.
This feature selection mechanism provides robustness against over-fitting issues. We configure random forest as an ensemble of 100 decision trees with each decision tree trained using $int(\log M + 1)$ features, where M is the total number of features.

\tool's random forest model classifies network requests based on the provenance (creation and modification history) of a node and the context around it.
These classification decisions are made before network request are sent, so
that \tool can prevent network communication with ad and tracking related
parties. 
A single node may initiate many network requests (either due to it being a
script node, or being modified by script to reference multiple
resources).
As a result, any node may be responsible for an arbitrary number of network requests. 
\tool classifies three categories of network requests:

\begin{enumerate}
    \item Requests initiated by the webpage's HTML (e.g. the image referenced by
          an \texttt{<img>} tag's \texttt{src} attribute).
    \item Requests initiated by a node's attribute change (e.g. a new background
          image being downloaded due to a new CSS style rule applying because of
          a mouse hover).
    \item Requests initiated directly by \JS code (e.g. \AJAX requests, image
          objects not inserted into the DOM).
\end{enumerate}

\vspace{.1in}
\section{\tool Evaluation}
\label{sec:eval}
In this section we evaluate the accuracy, usability, and performance of \tool when applied to live, real-world, popular websites.  


\subsection{Accuracy}
\label{sec:eval:ground-truth}
We first evaluate how accurately \tool is able to distinguish advertising and tracking content from benign web resources. 


\point{Ground Truth}
To evaluate \tool's accuracy, we first need to gather a ground truth to label a large number of ad/tracking related network requests.
We generate a trusted set of ground truth labels by combining popular crowdsourced filter lists that target advertising and/or tracking,
and applying them to popular websites.
Table~\ref{table:ground-truth} lists the \NumFilterLists popular filter lists we combine to form our ground truth.  
These lists collectively contain more than a hundred thousand crowdsourced rules for determining whether a URL serves advertising and/or tracking content.

\begin{table}[h]
    \centering
    \begin{tabular}{lrr}
        \toprule
            List                    & \# Rules  & Citation \\
        \midrule
            EasyList                & \GroundTruthNumRulesEasyList    & \cite{easylist_web} \\
            EasyPrivacy             & \GroundTruthNumRulesEasyPrivacy    & \cite{EasyPrivacy} \\
            Anti-Adblock Killer     & \GroundTruthNumRulesAntiAdblock     & \cite{killer} \\
            Warning Removal List    & \GroundTruthNumRulesWarningRemovalList       & \cite{warning} \\
            Blockzilla              & \GroundTruthNumRulesBlockzilla     & \cite{blockzilla} \\
            Fanboy Annoyances List  & \GroundTruthNumRulesFanboy    & \cite{fanboy_annoynace} \\
            Peter Lowe's List       & \GroundTruthNumRulesPeterLowe     & \cite{peter_lowe_list} \\
            Squid Blacklist         & \GroundTruthNumRulesSquid & \cite{squid_blacklist} \\
        \bottomrule
        \\
    \end{tabular}
    \caption{Crowd sourced filter lists used as ground truth for identifying ad
             and tracking resources. Rule counts are as of Nov. 12, 2018.}
    \label{table:ground-truth}
\end{table}

Advertising resources include audio-visual promotional content on a website.
Tracking resources collect unique identifiers (e.g., cookies) and sensitive information (e.g., browsing history) about users.
%
In practice, there is no clear division between ad and tracking resources.  
Many resources on the web not only serve advertising images and videos but also track the users who view it. 
It is also noteworthy that EasyList (to block ads) and EasyPrivacy (to block trackers) have a significant overlap.
Because of this overlap, we do not attempt to distinguish between advertising and tracking resources.

Note that while these crowdsourced filter lists suffer from well-known shortcomings \cite{Vester2018WhoFilterstheFiltersarXiv}, 
we treat them as ``trusted'' for three reasons. 
First, they are reasonably accurate for top-ranked websites 
even though they suffer on low-ranked websites \cite{Merzdovnik17BlockMeIfYouCanESP,Englehardt16MillionSiteMeasurementCCS}.
Second, a more accurate alternative, building a web-scale, manually generated, 
expert set of labels would require labor and resources far beyond what is feasible for a research project. 
Third, we use several filter lists together to maximize their coverage and reduce false negatives.

We visit the homepages of the Alexa top-10K websites with our instrumented Chromium browser.
We expect that the top-10K websites is a diverse and
large enough set to contain most common browsing behaviors.
We limit our sample of websites to the 10K most popular sites to
avoid biasing our sample; previous work has found that popular filter lists
work reasonably well for popular sites~\cite{Merzdovnik17BlockMeIfYouCanESP,Englehardt16MillionSiteMeasurementCCS}.
Applying crowdsourced filter lists to unpopular sites (sites that, almost by definition,
the curators of filter lists are less likely to visit) risks skewing our
data set to include a large number of false negatives (i.e. advertising and
tracking resources that filter list authors have not encountered).

We apply filter lists to websites in the
following manner.  We visit the homepage of each site with our instrumented
version of Chromium and wait for each page to finish loading (or
\CrawlerDwellSecs seconds, whichever occurs first). Next we record every URL
of every resource fetched when loading and rendering each page. 
We then label each fetched resource URL as \AdLabel and \NoAdLabel, 
based on the whether they are identified as ad or tracking related by any of a set of filter lists.
Our final labeled dataset consists of \DataSetNumResources URLs, 
fetched from \DataSetDomainsSuccesfullyCrawled successfully crawled domains.\footnote{The success rate of about 90\% in our crawl is in line with 
those of previous studies \cite{Merzdovnik17BlockMeIfYouCanESP,Englehardt16MillionSiteMeasurementCCS}.}


\point{Results}
\begin{table*}[ht]
    \centering
    \begin{tabular}{lrrr|rrrrr}
        \toprule
            Resource                &
            \# Resources            &
            Blocked by Filter Lists &
            Blocked by \tool        &
            Precision               &
            Recall                  &
            FPR                  &
            FNR                  &
            Accuracy \\
        \midrule
            Image                  &
                \DataSetNumImgResources   &
                \DataSetNumImgResourcesAdFilterList &
                \DataSetNumImgResourcesAdAdGraph &
                \DataSetNumImgResourcesPrecision &
                \DataSetNumImgResourcesRecall &
                \DataSetNumImgResourcesFPR &
                \DataSetNumImgResourcesFNR &
                \DataSetNumImgResourcesAccuracy \\
            Script                  &
                \DataSetNumScriptResources   &
                \DataSetNumScriptResourcesAdFilterList &
                \DataSetNumScriptResourcesAdAdGraph &
                \DataSetNumScriptResourcesPrecision &
                \DataSetNumScriptResourcesRecall &
                \DataSetNumScriptResourcesFPR &
                \DataSetNumScriptResourcesFNR &
                \DataSetNumScriptResourcesAccuracy \\
            CSS                  &
                \DataSetNumCSSResources   &
                \DataSetNumCSSResourcesAdFilterList &
                \DataSetNumCSSResourcesAdAdGraph &
                \DataSetNumCSSResourcesPrecision &
                \DataSetNumCSSResourcesRecall &
                \DataSetNumCSSResourcesFPR &
                \DataSetNumCSSResourcesFNR &
                \DataSetNumCSSResourcesAccuracy \\
            AJAX                  &
                \DataSetNumAjaxResources   &
                \DataSetNumAjaxResourcesAdFilterList &
                \DataSetNumAjaxResourcesAdAdGraph &
                \DataSetNumAjaxResourcesPrecision &
                \DataSetNumAjaxResourcesRecall &
                \DataSetNumAjaxResourcesFPR &
                \DataSetNumAjaxResourcesFNR &
                \DataSetNumAjaxResourcesAccuracy \\
            iFrame                  &
                \DataSetNumIFrameResources   &
                \DataSetNumIFrameResourcesAdFilterList &
                \DataSetNumIFrameResourcesAdAdGraph &
                \DataSetNumIFrameResourcesPrecision &
                \DataSetNumIFrameResourcesRecall &
                \DataSetNumIFrameResourcesFPR &
                \DataSetNumIFrameResourcesFNR &
                \DataSetNumIFrameResourcesAccuracy \\           
            Video                  &
                \DataSetNumVideoResources   &
                \DataSetNumVideoResourcesAdFilterList &
                \DataSetNumVideoResourcesAdAdGraph &
                \DataSetNumVideoResourcesPrecision &
                \DataSetNumVideoResourcesRecall &
                \DataSetNumVideoResourcesFPR &
                \DataSetNumVideoResourcesFNR &
                \DataSetNumVideoResourcesAccuracy \\            
        \midrule
            Total &
                \DataSetNumResources &
                \DataSetNumResourcesAdFilterList &
                \DataSetNumResourcesAdAdGraph &
                \AdGraphPrecision &
                \AdGraphRecall &
                \AdGraphFPR &
                \AdGraphFNR &
                \DataSetNumResourcesAccuracy \\
        \bottomrule
        \\
    \end{tabular}
    \caption{Number of resources, broken out by type, encountered during our
             crawl, and incidence of ad and tracking content, as determined
             by popular filter lists and \tool.}
    \label{table:accuracy-by-resource}
    \vspace{-10pt}
\end{table*}
We use the random forest model to classify each fetched URL. We then
compare each predicted label with the label derived from our ground
truth data set, the set of filter lists described above. We then evaluate how
accurately our model can reproduce the filter list labels through a stratified
10-fold cross validation, and report the average accuracy. \tool classifies
\AdLabel and \NoAdLabel with a high degree of accuracy, achieving
\AdGraphAccuracy accuracy, with \AdGraphPrecision precision, and \AdGraphRecall
recall.

As Table~\ref{table:accuracy-by-resource} shows, \tool classifies web resources
with a high degree of accuracy. We note that \tool is more accurate in
classifying visual resources such as images (\DataSetNumImgResourcesAccuracy
accuracy) and CSS (\DataSetNumCSSResourcesAccuracy accuracy) than invisible
resources like \JS (\DataSetNumScriptResourcesAccuracy accuracy) and \AJAX
requests (\DataSetNumAjaxResourcesAccuracy accuracy). This suggests an interesting possibility,
that \tool's labels are correct, and filter lists miss-classify invisible resources due to their reliance on
human crowdsourced feedback. We investigate this possibility, and more broadly the causes of
disagreements between \tool and filter lists in the next subsection.

\subsection{Disagreements Between \tool and Filter Lists}
\label{subsec:disagreement}
We now manually analyze the cases where \tool disagrees with filter lists
to determine which labeling is incorrect, \tool's or filter lists'.
Overall, we find that \tool is able to identify many advertising and tracking resources 
missed by filter lists.  We also find that \tool correctly identifies many resources as benign which 
filter lists incorrectly block.
%
These findings imply that \tool's actual accuracy is higher than \AdGraphAccuracy.

\point{Methodology}
To understand why \tool disagrees with existing filter lists,  we perform a manual analysis of a sample of network requests where \tool identifies a resource as ad/tracking related but filter lists identify as benign (i.e. false positives) and where filter lists identify a resource as ad/tracking related but \tool identifies as benign (i.e. false negatives).
We select these ``false positives'' and ``false negatives'' from the  most frequent advertising and tracking related resource types: \JS code units and images.
We manually analyze all of the \FalsePositiveNumSuccessImageCases distinct images and a random sample of \FalsePositiveNumScriptCases script URLs that \tool classifies as \AdLabel but filter lists label as \NoAdLabel and a random sample \FalseNegativeNumImageCases  images and \FalseNegativeNumScriptCases script URLs that \tool classifies as \NoAdLabel but filter lists label as \AdLabel.
The goal of our manual analysis is to assign each \JS unit or image to one of the following labels:

\begin{enumerate}
      \item \textbf{True Positive}: \tool's classification is correct and the filter
        lists are incorrect; the resource is related to advertising or tracking.
      \item \textbf{False Positive}: The label by filter lists is correct and \tool's
        classification is incorrect; the resource is not related to
        advertising or tracking.
      \item \textbf{True Negative}: \tool's classification is correct and the filter
        lists are incorrect; the resource is not related to advertising or tracking.
      \item \textbf{False Negative}: The label by filter lists is correct and \tool's
        classification is incorrect; the resource is related to
        advertising or tracking.        
      \item \textbf{Mixed}: The resource is dual purpose (i.e. both ad/tracker and
        benign). This label is only used for script resources.
      \item \textbf{Undecidable}: It was not possible to determine whether the
        resource is an ad/tracker.
\end{enumerate}

We decide whether an image was advertising or tracking related through the following three steps.
First, we label all tracking pixels ($1\times1$ sized images used to initiate a cookie 
or similar state-laden communication) as ``true positive'' 
if \tool classified it as \AdLabel and ``false negative'' if \tool classified it as \NoAdLabel.
Second, we consider the content of each image and look for text indicating advertising, 
such as the word ``sponsored", prices, or mentions of marketers. 
If the image has such text, we consider the image as an advertisement 
and label it ``true positive'' if \tool classified it as \AdLabel and ``false negative'' 
if \tool classified it as \NoAdLabel.
If the case is ambiguous, such as an image of a product that could either be advertising 
or a third-party discussion of the product, we use the ``undecidable'' label.
Third, we label all remaining cases as ``false positive'' if \tool classified them as \AdLabel 
and ``true negative'' if \tool classified them as \NoAdLabel.

Deciding the labels for the sampled script resources is more challenging.
Determining the purpose of a \JS file requires inspecting and understanding large amounts of code,
most of which has no documentation, and which is in many cases minified or obfuscated. 
We label a script as ``true positive'' (advertising or tracking related) if most of the script performs any of the following functionality: 
cookie transmission, passive device fingerprinting, communication with known ad or tracking services, sending
beacons, or modifying DOM elements whose attributes are highly indicative of an
ad (e.g. creating an image carousel with the \texttt{id} ``ad-carousel'');
and \tool classified it as \AdLabel and ``false negative'' if \tool classified it as \NoAdLabel.
If the script primarily includes functionality distinct from the above (e.g. form
validation, non-ad-related DOM modification, first-party \AJAX server
communication), we label it as ``false positive'' if \tool classified it as \AdLabel 
and ``true negative'' if \tool classified it as \NoAdLabel.
If the script contains significant amounts of both categories of functionality, 
we label the script as ``mixed''. 
In cases where the functionality is not discernable, we use the ``undecidable'' label.

\point{False Positive Analysis}
Table~\ref{table:false-positive-analysis} presents the results of our disagreement analysis for false positives.
In cases where \tool identifies a resource as suspect, and filter lists label it as benign, 
\tool's determination is correct \FalsePositiveJSLowPct--\FalsePositiveJSHighPct of the time 
for \JS and \FalsePositiveImgPct of the time for images.

\tool is often able to detect advertising and tracking resources that are missed by filter lists. 
For example, \tool blocks a 1x1 pixel on \url{cbs.com} that includes a tracking identifier in its query string.
In another example, \tool blocks a script (\url{js1}) on \url{nikkan-gendai.com} that performs browser fingerprinting. 
Filter lists likely missed these resources because they are often slow to catch up when websites 
introduce changes \cite{Iqbal17AntiABIMC}.

There are however several false positives that are actual mistakes by \tool.
For example, \tool blocks a third-party dual purpose script (\url{avcplayer.js}), a video player library that also serves ads,
on \url{inquirer.net}.
%
Interestingly, \tool detects many such dual-purposed scripts that are beyond the ability of binary-label filter lists.

These results demonstrate that \tool is able to identify many edge case resources (e.g. mixed-use) 
that can be used to refine future versions of \tool.
As discussed in Section \ref{point:future improvements:ClassificationGranularity}, 
\tool can be extended to handle such mistakes by implementing more fine-grained blocking.

\begin{table}[htbp]
    \centering
    \begin{tabular}{l|rrrr}
        \toprule
            &
                \multicolumn{2}{c}{Image} &
                \multicolumn{2}{c}{Script} \\
            &
                \# & \% &
                \# & \% \\
        \midrule
            True Positive &
                \FalsePositiveImgTruePositiveNum & \FalsePositiveImgTruePositivePct &
                \FalsePositiveJSTruePositiveNum & \FalsePositiveJSTruePositivePct \\
            False Positive &
                \FalsePositiveImgFalsePositiveNum & \FalsePositiveImgFalsePositivePct &
                \FalsePositiveJSFalsePositiveNum & \FalsePositiveJSFalsePositivePct \\
            Mixed &
                \FalsePositiveImgSMixedNum & \FalsePositiveImgMixedPct &
                \FalsePositiveJSMixedNum & \FalsePositiveJSMixedPct \\
            Undecidable &
                \FalsePositiveImgUndecideableNum & \FalsePositiveImgUndecideablePct &
                \FalsePositiveJSUndecideableNum & \FalsePositiveJSUndecideablePct \\
        \bottomrule
        \multicolumn{5}{l}{}
    \end{tabular}
    \caption{Results of manual analysis of a sample of cases where \tool classifies a resource as \AdLabel and filter lists label it as \NoAdLabel.}
    \label{table:false-positive-analysis}
    \vspace{-10pt}
\end{table}

\point{False Negative Analysis}
Table~\ref{table:false-negative-analysis} presents the results of our disagreement analysis for false negatives.
In cases where \tool identifies a resource as benign, and filter lists label it as suspect, 
\tool's determination is correct \FalseNegativeJSLowPct--\FalseNegativeJSHighPct of the time 
for \JS and \FalseNegativeImgPct of the time for images.

Again, \tool is often able to identify benign content that is incorrectly over-blocked by filter lists. 
For example, \tool does not block \url{histats.com} when visited as a first-party in our crawl, 
but this domain is blanketly blocked by the Blockzilla filter list even when visited as a first-party.
In another example, \tool does not block a social media icon \url{facebook-gray.svg} 
(served on \url{postimees.ee} as a first-party resource) and a 
privacy-preserving analytics script \url{piwik.js} (served on \url{futbol24.com} as a first-party resource). 
It can be argued that many of these resources are neither ads nor pose a tracking threat 
\cite{KontaxisPrivacyPreservingSocialPlugins12Usenix, eff_facebook_letter}. 
Filter lists over-block in such cases because of the inclusion of overly broad
rules (e.g. blocking entire domains, or any URL containing a given string). 

There are however several false negatives that are actual mistakes by \tool. 
For example, \tool misses \url{fingerprint2.min.js} served by a CDN \url{cloudflare.com} on \url{index.hr}.
%
%
\tool likely made this mistake because a popular third-party CDN, which is typically used to serve functional 
content, is used to serve a fingerprinting script.
As discussed in Section \ref{point:future improvements:Features}, 
\tool can be extended to handle such mistakes by extracting new features from \JS APIs.

\begin{table}[htbp]
    \centering
    \begin{tabular}{l|rrrr}
        \toprule
            &
                \multicolumn{2}{c}{Image} &
                \multicolumn{2}{c}{Script} \\
            &
                \# & \% &
                \# & \% \\
        \midrule
            True Negative &
                \FalseNegativeImgTrueNegativeNum & \FalseNegativeImgTrueNegativePct &
                \FalseNegativeJSTrueNegativeNum & \FalseNegativeJSTrueNegativePct \\
            False Negative &
                \FalseNegativeImgFalseNegativeNum & \FalseNegativeImgFalseNegativePct &
                \FalseNegativeJSFalseNegativeNum & \FalseNegativeJSFalseNegativePct \\
            Mixed &
                \FalseNegativeImgSMixedNum & \FalseNegativeImgMixedPct &
                \FalseNegativeJSMixedNum & \FalseNegativeJSMixedPct \\
            Undecidable &
                \FalseNegativeImgUndecideableNum & \FalseNegativeImgUndecideablePct &
                \FalseNegativeJSUndecideableNum & \FalseNegativeJSUndecideablePct \\
        \bottomrule
        \multicolumn{5}{l}{}
    \end{tabular}
    \caption{Results of manual analysis of a sample of cases where \tool classifies a resource as \NoAdLabel and filter lists label it as \AdLabel.}
    \label{table:false-negative-analysis}
    \vspace{-10pt}
\end{table}

\subsection{Site Breakage}
\label{sec:eval:breakage}
Content blocking tools carry the risk of breaking benign site
functionality.  Content blockers prevent resources that the website expects to be in
place from being retrieved, which can have the carry over effect of harming desireable
site functionality, especially when tools mistakenly block benign resources~\cite{AdblockCauseBreakage}.
Thus assessing the usefulness of a content blocking approach must also include an
evaluation of how many sites are ``broken'' by the intervention.  

Next we evaluate how often, and to what degree, \tool breaks
benign (i.e. user desired) website functionality.  We do so by having two
human reviewers visit a sample of popular websites using \tool, and having
them independently record their assessment of whether the site worked correctly.
We find that \tool only affects benign functionality on a small number of sites,
and at a rate equal to or less than popular filter lists.

\point{Methodology}
We estimate how many sites \tool breaks by having two evaluators 
use \tool on a sample of popular websites and independently record their
determination of how \tool impacts the site's functionality.
Because of the time consuming nature of the task, we select a smaller sample of
sites for this breakage evaluation than we use for the accuracy evaluation. 

Our evaluators use \tool on two sets of websites: first the Alexa top-10 websites, 
and second on a random sample of 100 websites from the Alexa top-1K list,
resulting in a total of \TotalNumOfDataSet sites for breakage evaluation.

Automatic site breakage assessment is challenging due to the complexity of
modern web applications \cite{NikiforakisPriVaricator2015WWW,Yu16TrackingTheTrackersWWW}. Unfortunately,
manual inspection for site breakage assessment is not only time-consuming but
also likely to lose completeness as the functionalities of a website are often
triggered by certain events that may be hard to manually cover exhaustively. 
As a tradeoff, we adopt the approach from \cite{snyder2017most}, which is a
manual analysis but focuses on \textit{the user's perspective}. In other words,
we intentionally ignore the breakages that only affect the website owner as
they do not have any impact on user experience.

For each website, our evaluators independently perform the following steps.

\begin{enumerate}
	\item Open the website with stock Chromium, as a control, and perform as many actions as possible within two
        minutes. We instruct our evaluators to exercise the kinds of behaviors
        that would be common on each site.  For example, in a news site this might be browsing through an article; on a
        e-commerce site this might include searching for a product and proceeding
        to checkout etc.
	\item Open the website with \tool, repeat the actions performed above, and
        assign a breakage level of
        \begin{enumerate}[label=(\alph*)]
            \item \textbf{no breakage} if there is no perceptible difference between
                  \tool and stock Chromium;
            \item \textbf{minor breakage} if the browsing experience is altered, but
                  objective of the visit can still be completed; or
            \item \textbf{major breakage} if objective of the visit cannot be
            completed.
        \end{enumerate}
      \item Open the website with Adblock Plus\footnote{Adblock Plus is
      configured with the same \NumFilterLists filter lists that are used to train \tool.}, repeat the actions, and assign a breakage level as above.
\end{enumerate}
To account for the subjective nature of this analysis, we have each evaluator
visit the same sites, at similar times, and determine their ``breakage'' scores independently. Our
evaluators give the same score \AgreementRatio of the time, supporting the
significance of their analysis.

\begin{table}[ht]
    \centering
    \begin{tabular}{l|rrrrrrrr}
        \toprule
            Tool & \multicolumn{2}{c}{No breakage} & \multicolumn{2}{c}{Major} & \multicolumn{2}{c}{Minor} & \multicolumn{2}{c}{Crash} \\
                 & \# & \% & \# & \% & \# & \% & \# & \% \\ 
        \midrule
            \tool & \AGNoBreakageNum & \AGNoBreakagePct & \AGMajorBreakageNum & \AGMajorBreakagePct & \AGMinorBreakageNum & \AGMinorBreakagePct & \AGCrashBreakageNum & \AGCrashBreakagePct \\
        Filter lists & \FLNoBreakageNum & \FLNoBreakagePct & \FLMajorBreakageNum & \FLMajorBreakagePct & \FLMinorBreakageNum & \FLMinorBreakagePct & \FLCrashBreakageNum & \FLCrashBreakagePct \\
        \bottomrule
        \multicolumn{9}{l}{}
    \end{tabular}
    \caption{Breakdown of breakage analysis results (\# columns are the average of two independent scores.)}
    \label{table:breakage_breakdown}
\end{table}

\point{Results}
Table \ref{table:breakage_breakdown} reports the site breakage assessments as
the average of two reviewers. The evaluation shows that \tool and filter lists are
comparable in terms of site breakage. \tool and filter lists do not cause any
breakage on \AGNoBreakagePct and \FLNoBreakagePct of the sites, respectively.
The major breakage rate (\AGMajorBreakagePct) is also on par with the filter
lists (\FLMajorBreakagePct).
We also note that \tool's breakage is much lower than other commonly used
privacy oriented browsers (e.g. 16.3\% for Tor Browser \cite{snyder2017most}).

\subsection{Feature Analysis}
\label{sec:eval:ffeatureanalysis}
Next, we discuss the intuitions behind some of the features used in \tool,
and evaluate their ability to distinguish ad/tracking content
from benign content. We describe some of the
features that are most useful (in terms of information gain\cite{QuinlanInformationGain})
in \tool's predictions.

\point{Structural Features}
Two of the structural features that provided the highest information gain are a node's \textit{average degree connectivity} and its \textit{parents' attributes}.

\begin{figure}[!htpb]
      \centering
          \subfigure[average degree connectivity]{
          \includegraphics[width=0.45\columnwidth]{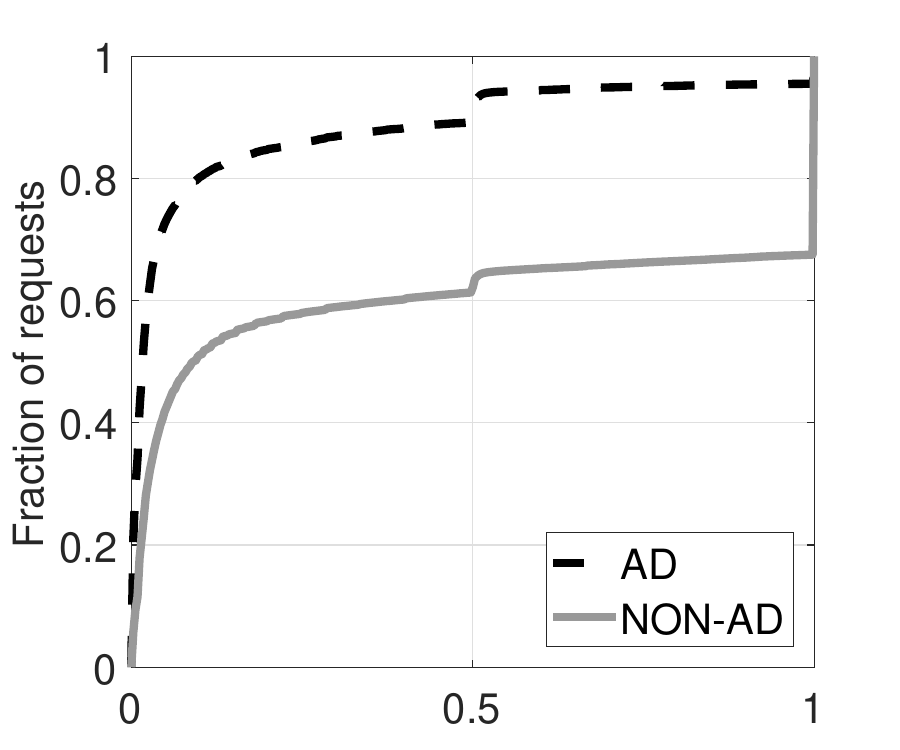}
          \label{figure:avg_degree_connectivity}
          } \subfigure[async script]{
          \includegraphics[width=0.45\columnwidth]{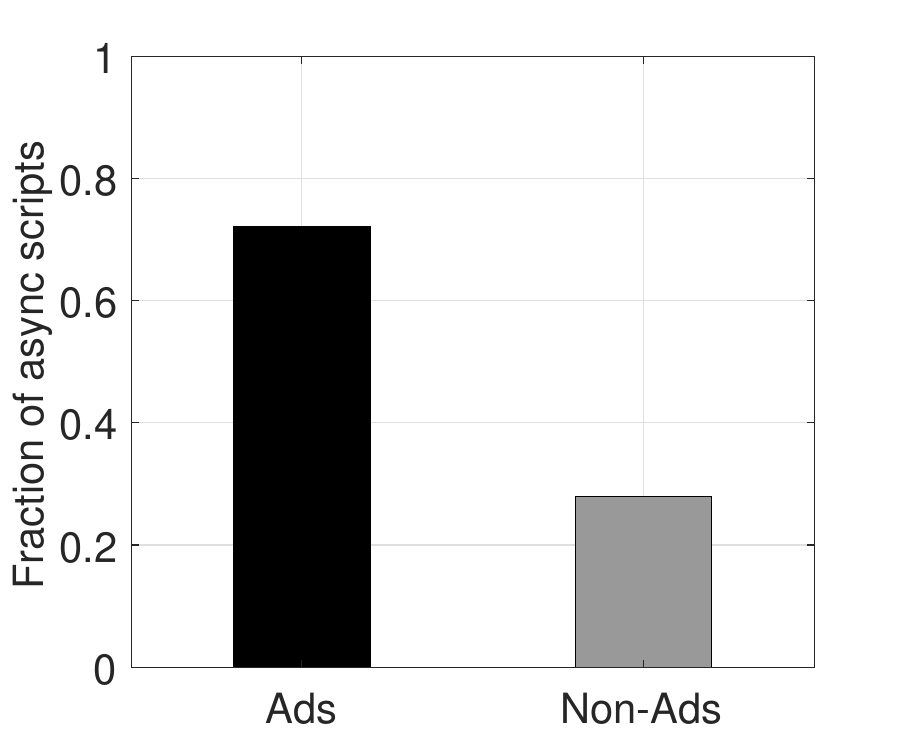}
          \label{figure:first_parent_async}
          }  
      \caption{Conditional distributions for structural features.}
      \label{figure:structural_feature_distributions}
  \end{figure}

We expect \AdLabel nodes to have lower \textit{average degree connectivity},
since the interaction of these nodes is confined to only ad/tracking content, and thus
appear in less connected cliques.  Conversely, we expect that \NoAdLabel nodes appear
alongside, and interact with, functional content more, and thus have
higher \textit{average degree connectivity}. Our results in Figure
\ref{figure:avg_degree_connectivity} support this
intuition.  \AdLabel nodes do indeed have have lower \textit{average degree connectivity}
than \NoAdLabel nodes.

We also expect the parents of \AdLabel nodes to have different
attributes than \NoAdLabel nodes.  This intuition came from the expectation
that \AdLabel nodes are more likely to follow common practices and
standards, such as those proposed by the Interactive Advertising Bureau (IAB)\cite{iab_web}.
For example, IAB's LEAN standard \cite{iab_lean} requires ad related scripts to
load asynchronously (indicated by the presence of the \texttt{async} attribute
on a script node).  We capture this intuition in a feature by considering
the attributes of each network requests' parent nodes (in our graph
representation, the parent of a network request might be the script
element that initiates the network request). Our results in Figure \ref{figure:first_parent_async} support this intuition. 
The parents of \AdLabel nodes with \texttt{script} tag name were 3 times
more likely to have the \texttt{async} attribute than \NoAdLabel nodes.

We note that some structural features are more robust to obfuscation than others.
For example, to flummox the classifier, it would be more challenging for an adversary to manipulate 
a node's average degree connectivity (which depends on all of the node's neighbors)
than it would be to manipulate the attributes of a parent node.

\point{Content Features}
Two of the content features that provided the highest information gain are a node's \textit{domain party} and its \textit{URL length}.

\begin{figure}[!htpb]
    \centering
 \subfigure[domain party]{
        \includegraphics[width=0.45\columnwidth]{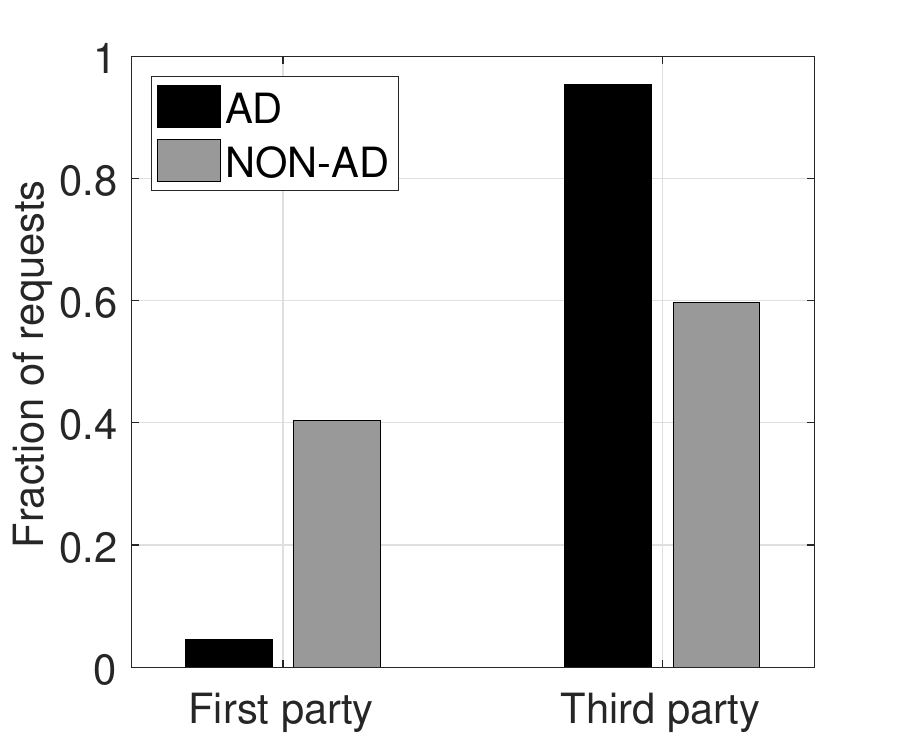}
        \label{figure:domain_party}
        }
        \subfigure[length of URL]{
        \includegraphics[width=0.45\columnwidth]{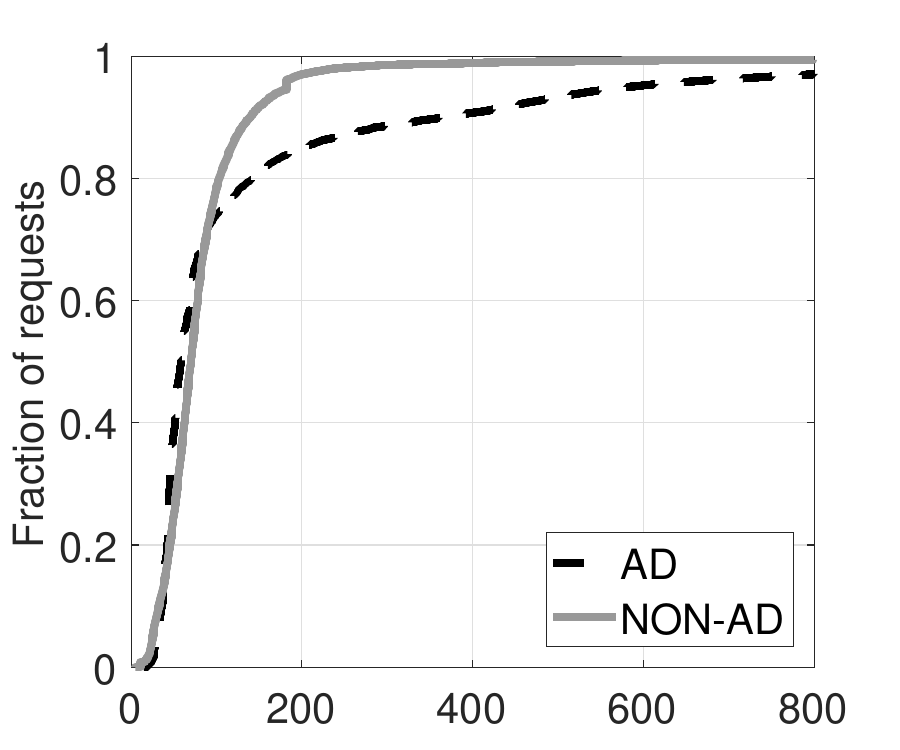}
        \label{figure:url_length}
        }
        \caption{Conditional distributions for content features.}
    \label{figure:feature_distributions}
\end{figure}

We expect \AdLabel nodes to be more likely to come from third-party
domains than \NoAdLabel nodes.  We capture this intuition in a boolean feature,
recording whether the domain of a network request differs from domain of the first
party document.  Figure \ref{figure:domain_party} shows this intuition to be correct.
More than 90\% of the ads came from third-party domains.

We also expect \AdLabel nodes to include a large number of query
parameters in their URLs.
We capture this intuition by using a request's \textit{URL length} as a numeric feature. 
Figure \ref{figure:url_length} shows this intuition to be correct. 
\AdLabel node URLs were on average longer than \NoAdLabel node URLs.

We again note that some content features are more robust to obfuscation than others.
For example, to flummox the classifier, it would be more challenging for an adversary to switch ads/trackers 
from third-party to first-party that it would be to manipulate the length
of a URL.

\point{Ablation Analysis}
Next, we separately evaluate structural and content features in terms of their contribution to \tool's accuracy.
To this end, we train additional classifiers separately, one using only structural features, the other using only content features.
While structural features and content features have comparable accuracy 
they provide complementary information, which when used together improve \tool's accuracy.
For example, excluding structural features results in a decrease of 
\PrecisionDropWithoutStructuralFeatures in precision, 
\RecallDropWithoutStructuralFeatures in recall, and \AccuracyDropWithoutStructuralFeatures in accuracy.

We also expect structural features to be more robust than content features.
Structural features consider neighboring graph structure of a node while content features only consider a node in isolation.
To manipulate the structural features, an adversary would need to change the target node, its neighbors, 
and subsequently their neighbors.
Manipulating content features would only require changing the target node.

Thus, we conclude that the graph-based representation of \tool, as captured by structural features, 
contributes to its accuracy and robustness.

\subsection{Tradeoffs in Browser Instrumentation}
Recall from Section \ref{sec:classifier:construction} that \tool modifies Chromium to attribute DOM modifications to \JS code units. 
This is different from most existing content blocking tools that operate at the extension layer.
\tool's browser instrumentation is a trade off; it gains
attribution accuracy at the cost of ease of distribution. 
This raises the question of whether \tool can instead be implemented as a browser extension
on any web browser.

We investigate this question by implementing \tool as
a browser extension, using the best possible attribution option available
at the extension layer (\JS stack walking, discussed in
Section~\ref{sec:background:attribution}). 
We test the accuracy of the best possible extension implementation of \tool by re-crawling 
the Alexa-10k with a modified version of \tool, using the same
methodology described in Section~\ref{sec:eval:ground-truth}.  
This modified version of \tool uses \JS stack walking to attribute DOM
modifications to script units, instead of the Blink and \VJS modifications.  
We then train and test ML classifier on the graphs constructed using \JS stack walking.

We compare the accuracy of
this best-possible-extension implementation to our in-browser 
implementation of \tool. We find that implementing \tool as a browser
extension significantly reduces classification accuracy. 
Implementing \tool as a browser extension degrades precision by
\AdGraphStkWalkPrecisonImprv, recall by \AdGraphStkWalkRecallImprv, and
accuracy by \AdGraphStkWalkAccuracyImprv.  
Thus, the mistakes \JS stack walking makes in attribution
lead to more errors in classification.  
We conclude that costs of implementing \tool's as a set of browser modifications
(i.e. difficulty in distribution) is more than offset by the benefits (i.e.
increased classification accuracy), and that \tool is best implemented
as Blink and V8 modifications.

\subsection{Performance}
\label{sec:eval:performance}
We evaluate \tool's performance as compared to stock Chromium and Adblock Plus. 
\tool performs faster in most cases than the most popular blocking tool, Adblock Plus, and in many cases results in
faster performance than stock Chromium. 
This is the result of both careful engineering in \tool's implementation, and
\tool's instrumentation overhead (often) being more than offset
by the network and rendering savings gained by having to fetch and render less
page content (i.e. the content blocked by \tool).

\begin{figure}
  \centering
   \vspace{-15pt}
  \includegraphics[width=.4\textwidth]{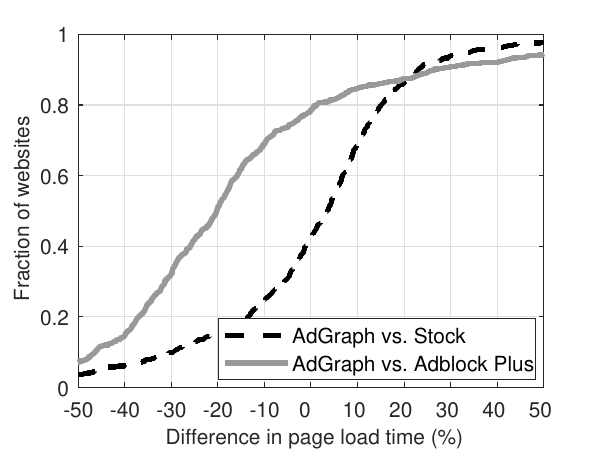}  
  \caption{Overhead ratio in terms of page load time.}
  \vspace{-15pt}
  \label{figure:page_load_time_overhead_ratio}
\end{figure}

To measure whether \tool is a practical blocking solution, we compare the performance
of \tool, stock Chromium, and Chromium with Adblock Plus
installed (using Adblock Plus's default configuration) on the Alexa 1K.  Our simulated network uses a 10 Mbps
downlink with a latency of 100ms.  We visit the landing page of each website 10 times and record the average page load time (measured 
as the difference between the DOM's \texttt{navigationStart} and
\texttt{loadEventEnd} events).
Figure~\ref{figure:page_load_time_overhead_ratio} presents \tool's page load time 
compared to stock Chromium, and Chromium with Adblock Plus.$^{6}$
\begin{table}[t!]
  \centering
  \begin{tabular}{lrr}
      \toprule
          Resource Type     & \tool faster           & Chromium faster \\
      \midrule
          Image              & \AGStockImageBetter    & \AGStockImageWorse \\
          Script             & \AGStockScriptBetter   & \AGStockScriptWorse \\          
          CSS                & \AGStockCSSBetter      & \AGStockCSSWorse \\
          AJAX               & \AGStockAJAXBetter     & \AGStockAJAXWorse \\
          iFrame             & \AGStockIframeBetter   & \AGStockIframeWorse \\                              
          Video              & \AGStockVideoBetter    & \AGStockVideoWorse \\
      \bottomrule
      \\
  \end{tabular}
  \caption{Comparison of average percentage of resources \tool blocks on sites
           where \tool outperforms Chromium, and vise versa.  For
           all resource types, \tool performs faster when more resources are blocked.}
  \label{table:perf-diff}
\end{table}

\tool performs faster than Chromium on \PerfAvgVsChrome of websites.  
\tool is often faster than stock Chromium because it needs 
to fetch and render fewer resources than stock Chromium (i.e. the network requests
blocked by \tool). 
Table~\ref{table:perf-diff} shows that \tool outperforms Chromium on sites where it blocks more ad/tracking content, 
as compared to sites where it blocks less. 
Put differently, the more content \tool blocks, the
more it is able to make up for the instrumentation and classification overhead
with network and rendering savings.

\tool performs faster than Adblock Plus on \PerfAvgVsAdBlock of websites.
\tool is faster than Adblock Plus for two reasons.
First, Adblock Plus implements element hiding rules (i.e. rules describing elements that are still
fetched, but hidden when rendering), which carries with it an enforcement and
display-reflow overhead \tool does not share.
Second, \tool's blocking logic is implemented in-browser which leads
to performance improvement over Adblock Plus's implementation at the extension layer.

Overall, we conclude that \tool is performant enough to be a practical online content blocking solution.  
Future implementation refinements, and the exploration of cheaper
features, could further improve \tool's performance.

\vspace{.1in}
\section{Discussions}
\label{sec:discussion}

\subsection{Offline Application of \tool}
\label{subsection: offline adgraph}
\tool is designed and implemented to be used as an online, in-browser blocking tool. 
This is different than most blocking tools, which operate as extensions on mainstream browsers (e.g. Chrome, Firefox).
Since \tool requires browser instrumentation, it cannot be directly used by extension-based blockers that rely on offline manually curated filter lists. 
\tool can benefit existing blocking tools through the creation and maintenance of filter lists in several ways.

First, the accuracy of filter lists suffers because of they are manual generated and rely on informal crowdsourced feedback.
As discussed in Section \ref{subsec:disagreement}, filter list maintainers can 
analyze disagreements between \tool and filter lists to identify and fix potential inaccuracies in filter lists

Second, \tool can support the generation of filter lists targeting
under-served languages or region on the web.
Filter lists are inherently skewed towards popular websites and languages because of their larger and more active blocking user base \cite{Englehardt16MillionSiteMeasurementCCS,Merzdovnik17BlockMeIfYouCanESP}. 
Filter list maintainers receive much less feedback to fix inaccuracies on less popular websites.
This makes the creation and maintenance of filter lists for underserved regions
(geographically and linguistically) difficult, since these sites have less visitors.
Language/region specific filter lists are updated much less frequently than general (and mostly English targeting)
filter lists like EasyList.
Many languages and regions (most notably Africa) do not have dedicated filter lists at all.
\tool can assist in automatically generating filter lists for smaller or underserved regions.

Third, the manual nature of filter list maintenance has lead to increasing number of outdated and stale rules. 
Filter list rules can quickly get outdated because most websites frequently update and are highly dynamic.
Prior research found that filter lists can take months to update in response to such changes \cite{Iqbal17AntiABIMC}. 
Even when filter lists are updated, new rules are typically added (rather than editing old rules) which leads to accumulation of stale rules over time.  
Prior research reported that only 200 rules account for 90\% blocking activity for EasyList \cite{Vester2018WhoFilterstheFiltersarXiv}.
In other words, the number of rare-to-never used rules in EasyList is increasing over time which has performance implications. 
\tool can by used by filter list maintainers to periodically audit filter lists for identifying outdated and stale rules. 


\subsection{\tool Limitations And Future Improvements}
\label{subsection: future improvements}

\point{Ground Truth}
\tool relies on filter lists as ground truth to train a ML classifier for detecting ads/trackers. 
As we showed in Sections \ref{subsec:disagreement}, filter lists suffer from inaccuracies  
due to both false negatives and false positives.
\tool can address these inaccuracies in ground truth by gathering valuable user feedback 
when it is deployed at scale.
\tool can retrain its ML classifier periodically on improved ground truth as user feedback is received.

\point{Features}
\label{point:future improvements:Features}
The features used by \tool are manually designed, based on our domain knowledge and expert intuition, 
with the goal of achieving decent accuracy.
Note that the feature set is by no means ``complete'' and there is room for additional feature engineering 
to further improve accuracy.
New features can be systematically discovered by incorporating user feedback, 
which may reveal new characteristics of ads/trackers over time that are not currently covered by \tool.
%
%
New features may require addition of new instrumentation points such as \JS APIs 
or new feature modalities altogether, such as image based perceptual information \cite{grant17futureadblocking,SentinelAdblock,Tramer2018AdversarialarXiv}.
%

\point{Classification Granularity}
\label{point:future improvements:ClassificationGranularity}
\tool is currently designed to make binary decisions to either block or allow network requests.
However, as discussed in Section~\ref{subsec:disagreement}, \tool is also able to detect cases when a single \JS is used for both ad/tracking and functional content.
The cases where \JS code serves dual-purpose are challenging because blocking the request may break page functionality, while allowing the request will allow ads/trackers on the page.  
\tool's context rich classification approach can be adapted to more than two labels for handling such dual-purpose scripts.
Specifically, \tool can be trained at a more granular level to distinguish between ads/trackers, functional, and dual-purpose resources. 
\tool can respond to such dual-purpose resources with different remediations than outright allowing/blocking, such as giving those scripts a reduce set of DOM capabilities (e.g. reading/writing cookies \cite{firefox_StorageAccessPolicy,safari_itp}, access to certain APIs \cite{Brave_FingerprintingProtectionMode,TOR_FIngerprintProtection}), or blocking network requests issued from such scripts.

\vspace{.1in}
\section{Conclusion}
\label{sec:conclusion}
In this paper we proposed \tool, a graph-based ML approach to ad and tracker blocking.
We designed \tool to leverage fine-grained interactions between network requests, DOM elements, and JavaScript code execution to construct a graph representation that is used to trace relationships between ads/trackers and the rest of the page content.
To implement \tool, we instrumented Chromium's rendering engine (Blink) and JavaScript execution engine (\VJS) to efficiently gather complete HTML, HTTP, and JavaScript information during page load.
We leveraged this rich context by extracting distinguishing features to train a ML classifier for in-browser ad and tracker blocking at runtime.

We showed that \tool not only blocks ads/trackers with \AdGraphAccuracy accuracy but uncovers many ad/tracker and functional resources that are missed and over-blocked by filter lists, respectively.
We also showed that \tool's breakage is on par with  filter lists.
In addition to high accuracy and comparable breakage, we showed that \tool loads pages much faster as compared to existing content blocking tools.

We designed \tool to be used both online (for in-browser blocking) and offline (filter list curation).
Since the vast majority of extension-based blocking tools currently rely on manually curated filer lists, \tool's offline use case will aid filter list monitoring and maintenance. 
Overall, we believe that \tool significantly advances the state-of-the-art in ad and tracker blocking.

\section*{Acknowledgment}
This work is supported in part by the National Science Foundation under grant numbers 1715152, 1719147, and 1815131.

\balance
{\normalsize \bibliographystyle{acm}}
\bibliography{references}

\end{document}